\documentclass[prb,preprint,amsmath,amssymb,superscriptaddress,floatfix]{revtex4}
\usepackage{epsfig}
\usepackage{graphicx}
\usepackage{amsmath,amssymb}
\usepackage{xcolor}
\usepackage[export]{adjustbox}
\usepackage{bm}
\usepackage{multirow}

\newcommand\ddfrac[2]{\frac{\displaystyle #1}{\displaystyle #2}}
\def\ve{\varepsilon}

\def\unit#1{\ \mathrm{#1}}

\begin{document}

\title{Parity-controlled spin-wave excitations in synthetic antiferromagnets}

\author{A. Sud}
 \email{aakanksha.sud.17@ucl.ac.uk}
\affiliation{London Centre for Nanotechnology, University College London, 17-19 Gordon Street, London, WCH1 0AH, UK}
\author{Y. Koike}
 \affiliation{Department of Applied Physics, Tohoku University, Aoba 6-6-05, Sendai, 980-8579, Japan}
 \affiliation{WPI Advanced Institude for Materials Research, Tohoku University, 2-1-1, Katahira, Sendai 980-8577, Japan}
\author{S. Iihama}
 \affiliation{Frontier Research Institute for Interdisciplinary Sciences, Tohoku University, Sendai 980-8578, Japan}
  \affiliation{Center for Spintronics Research Network, Tohoku University, Sendai, 980-8577, Japan}
  \author{C. Zollitsch}
\affiliation{London Centre for Nanotechnology, University College London, 17-19 Gordon Street, London, WCH1 0AH, UK}
\author{S. Mizukami}
 \affiliation{WPI Advanced Institude for Materials Research, Tohoku University, 2-1-1, Katahira, Sendai 980-8577, Japan}
 \affiliation{Center for Spintronics Research Network, Tohoku University, Sendai, 980-8577, Japan}
 \affiliation{Center for Science and Innovation in Spintronics, Tohoku University, Sendai, 980-8577, Japan}
\author{H. Kurebayashi}%
 \email{h.kurebayashi@ucl.ac.uk}
 \affiliation{London Centre for Nanotechnology, University College London, 17-19 Gordon Street, London, WCH1 0AH, UK}
\affiliation{Dept.\ of Electronic \& Electrical Engineering, University College London, London WC1E 7JE, United Kingdom}

\maketitle
\textbf{We report in this study the current-induced-torque excitation of acoustic and optical modes in Ta/NiFe/Ru/NiFe/Ta synthetic antiferromagnet stacks grown on SiO$_2$/Si substrates. The two Ta layers serve as spin torque sources with the opposite polarisations both in spin currents and Oersted fields acting on their adjacent NiFe layers. This can create the odd symmetry of spatial spin torque distribution across the growth direction, allowing us to observe different spin-wave excitation efficiency from synthetic antiferromagnets excited by homogeneous torques. We analyse the torque symmetry by in-plane angular dependence of symmetric and anti-symmetric lineshape amplitudes for their resonance and confirm that the parallel (perpendicular) pumping nature for the acoustic (optical) modes in our devices, which is in stark difference from the modes excited by spatially homogeneous torques. We also present our macrospin model for this particular spin-torque excitation geometry, which excellently supports our experimental observation. Our results offer capability of controlling spin-wave excitations by local spin-torque sources and we can explore further spin-wave control schemes based on this concept.}

Synthetic antiferromagnets (SyAFs) are an excellent platform to explore novel spintronic and magnonic concepts with coupled magnetic moments\cite{DuineNatPhys2018,ChumakNatPhys2015}. Unlike a homogeneously magnetised single-layer ferromagnet, coupled magnetic layers are able to offer rich magnetic states where the competition between interlayer exchange coupling, external-field-induced Zeeman interaction as well as other magnetic anisotropy terms plays a role. This can expand into their dynamic regimes as the coupled moment nature inherently provides two eigenmodes, acoustic and optical modes\cite{Keffer1952,Krebs1990,Rezende_JAP2019}, where time-dependent components of two coupled moments are oscillating in-phase (acoustic) and out-of-phase (optical)in a spin-flop (canted) regime. These pure magnetic modes have been studied and discussed already around 1990s, e.g.~by Grunberg et al.\cite{GrunbergPRL1984} and Zhang et al.\cite{ZhangPRB1994}, followed by a number of more recent reports to investigate magnetic dynamics in SyAFs for various topics \cite{TaniguchiPRB2007,KonovalenkoPRB2009,SekiAPL2009,ChibaPRB2015,YangAPL2016,WangAPL2018,Kamimaki_APL2019,SorokinPRB2020,IshibashiSciAdv2020,ShiotaPRL2020,SudPRB2020}. SyAFs are also a subject of spin-orbit torque (SOT) excitations so far at (or close to) the dc limit\cite{BiPRB2017,KongNComm2019,ZhangPRB2018,MoriyamaPRL2018,IshikuroPRB2020,MasudaPRB2020} and there has been little in the literature on the use of SOTs for exciting SyAF spin-waves at GHz frequencies.

When we excite spin-waves by oscillating magnetic fields, the spatial symmetry/profile of microwave excitation determines which spin-wave modes are excited. In the simplest case, a uniform distribution of microwave excitation fields across a magnet can excite the uniform spin-wave mode (wavevetor $k$ = 0) as well as higher-order spin-wave resonance modes ($k$ $>$ 0) with odd index numbers since the spatial profile of the microwave excitation and spin-wave mode amplitude (with phase) matches to each other in terms of symmetry. Higher-order spin-wave resonance modes with even index numbers are excluded because the mode overlapping between microwaves and the spin-waves becomes zero when it is integrated over the real space. This in turn suggests that it should be possible to control the selection rules of spin-wave excitations by designing the spatial profile of excitation fields. Accessing hidden spin-wave states as well as tailoring spin-wave excitation efficiency by this approach has not been much explored in the past. 

In this $Letter$, we report our study of spin-wave excitation symmetry control by spatially anti-symmetric spin torques in our SyAF devices. By utilising local spin excitations due to spin-Hall effect (SHE) and Oersted fields from adjacent heavy-metal Ta layers, we control the spin-wave excitation nature in the SyAF devices. We observe the torque symmetry of parallel (perpendicular) pumping configuration of acoustic (optical) modes excited and measured in our devices, which is signature of the anti-symmetric spin torques for the two coupled moments. We provide analytical expressions for rectification voltages calculated based on the Landau-Lifshitz-Gilbert (LLG) equation for coupled magnetic moments at the macrospin limit. The model equations fully support the torque symmetry we observed in our experiments as well as allow to quantitatively analyse spin-orbit transport parameters such as the spin-Hall angle.

Before describing our experimental results, we here generalise the torque symmetry of optical and acoustic mode excitations ($\boldsymbol{\tau_\text{op}}$ and $\boldsymbol{\tau_\text{ac}}$). In SyAFs for in-plane field canted conditions, optical and acoustic modes can be generated by adding two individual magnetic moments by using $\pi$ rotation ($C_2$ operation) with respect to the applied magnetic field direction\cite{McNeillPRL2019}. We can combine the excitation terms for each moment by following the same manner and produce the torque expressions for both modes as:
\begin{equation}
\boldsymbol{\tau_\text{op}}=\boldsymbol{m_\text{1}^0}\times (\boldsymbol{B_\text{1}} + C_2\boldsymbol{B_\text{2}}) + \boldsymbol{m_\text{1}^0}\times \{\boldsymbol{m_\text{1}^0} \times (\boldsymbol{s_\text{1}}+C_2\boldsymbol{s_\text{2}})\};
\end{equation}
\begin{equation}
\boldsymbol{\tau_\text{ac}}=\boldsymbol{m_\text{1}^0}\times (\boldsymbol{B_\text{1}} - C_2\boldsymbol{B_\text{2}}) + \boldsymbol{m_\text{1}^0}\times \{\boldsymbol{m_\text{1}^0} \times (\boldsymbol{s_\text{1}}-C_2\boldsymbol{s_\text{2}} \}.
\end{equation}
Here, $m_\text{1}^0$, $\boldsymbol{B_\text{1}}$, $\boldsymbol{B_\text{2}}$, $\boldsymbol{s_\text{1}}$ and $\boldsymbol{s_\text{2}}$ are the time-independent component of magnetisation for one of the coupled moments, field-like torque for first and second magnetic moments and spin polarisation causing spin-transfer torques for first and second magnetic moments respectively. Full derivations of these two torque expressions are available in the Supplementary Material (SM). These expressions represent the torque symmetry of each mode excitation against the direction of external magnetic field ($\boldsymbol{B}$). When both moments are excited by uniform spin excitation, namely with the condition of $\boldsymbol{B_\text{1}}=\boldsymbol{B_\text{2}}$ and $\boldsymbol{s_\text{1}}=\boldsymbol{s_\text{2}}$, we can arrive at the following conclusions.
(i) When $\boldsymbol{B_\text{1}}$ and $\boldsymbol{s_\text{1}}$ are symmetric for the $C_2$ rotation, $\boldsymbol{\tau_\text{op}}$ is maximised ($\boldsymbol{B_\text{1}} + C_2\boldsymbol{B_\text{2}}=2\boldsymbol{B_\text{1}}$) and $\boldsymbol{\tau_\text{ac}}=0$, and (ii) when $\boldsymbol{B_\text{1}}$ and $\boldsymbol{s_\text{1}}$ are anti-symmetric for the $C_2$ rotation, $\boldsymbol{\tau_\text{op}} =0$ and $\boldsymbol{\tau_\text{ac}}$ is largest. Here the meaning of $\boldsymbol{B_\text{1}}$ and $\boldsymbol{s_\text{1}}$ being symmetric (anti-symmetric) for the $C_2$ rotation is the condition of $\boldsymbol{B_\text{1}} \| \boldsymbol{B}$ ($\boldsymbol{B_\text{1}} \perp \boldsymbol{B}$) which is in general termed as parallel (perpendicular) pumping configuration in magnetic dynamics. Altogether we can summarise that under a uniform excitation condition, the optical (acoustic) mode can be excited by parallel (perpendicular) pumping configuration. In the present study, we take this one step further to control the excitation symmetry by designing the local spin excitation configuration. When $\boldsymbol{m_1}$ and $\boldsymbol{m_2}$ experience non-uniform spin excitations, we find from Eqs. (1) and (2) that the perpendicular/parallel pumping nature can be tuned in our experiments. On the extreme case where $\boldsymbol{B_\text{1}}=-\boldsymbol{B_\text{2}}$ and $\boldsymbol{s_\text{1}}=-\boldsymbol{s_\text{2}}$, we  predict that the optical (acoustic) mode is excited by perpendicular (parallel) pumping configuration. This is because the odd-even parity of the spatial spin excitation is changed, leading to the parity change of excited spin-wave modes with this geometry. 

Films with SyAFs used in this study were prepared by magnetron co-sputtering techniques at a base pressure of 3$\times$10$\textsuperscript{-7}$ Pa. The films were grown on thermally oxidized Si substrates with stacking patterns of Ta(5)/NiFe(5)/Ru(0.4 or 0.5)/NiFe(5)/Ta(5) where the number in the brackets represents the thickness in nm. Prior to device fabrication, we characterised these films by vibrating magnetometry techniques to quantify the interlayer exchange coupling strength and to confirm the presence of spin-flop regimes in our films. From these films, rectangular bars with width (length) of 10 (40) $\mu$m were defined by standard photo-lithography techniques and Ar ion milling, prior to another Ti/Au bi-layer deposition for preparing a microwave waveguide on top of each bar. We exploit spin-transfer-torque ferromagnetic resonance (STT-FMR) techniques\cite{TulapurkarNature2005,LiuPRL2010} in order to excite spin-waves in our devices where the Ta layers are the source of spin torques acting on the adjacent NiFe layers individually. Crucially, the spin polarisation direction of spin torques in the Ta layers are opposite to each other due to the geometry, which achieves the anti-symmetric profile of these torques (e.g. $\boldsymbol{B_\text{1}}=-\boldsymbol{B_\text{2}}$) we utilise in our study. A schematic of our measurement set-up is shown in Fig.1(a). Vector magnets are used to generate $B$ at various in-plane angles $\phi$ with respect to current direction to map out the excitation symmetry in our devices. SHE\cite{SinovaRMP2015} in the two Ta layers convert electric currents into spin-currents injected into both magnetic layers where spin torques are exerted via the STT mechanism, together with field-like torques by Oersted fields. These cause a time-varying magnetisation precession at resonance, producing the time-varying resistance change due to anisotropic magnetoresistance (AMR). As a result, frequency mixing of two time-varying components (i.e. current and resistance) results in a time-independent voltage component we experimentally measure. As discussed later, we compare our experimental results with analytical solutions obtained from a macrospin model with the LLG equation for coupled magnetic moments.

Both acoustic and optical modes have been clearly identified for different excitation frequencies as shown in Fig. 1(b). The canted nature of synthetic antiferromagnets can be observed by the frequency dependence of resonance for both modes in our sample as shown in Fig. 1(c), indicating the presence of anti-ferromagnetic inter-layer exchange coupling through the Ru layer. The optical (acoustic) mode frequency becomes lower (higher) as $B$ is increased, as predicted by the following solutions of the LLG equation for coupled moments (See SM), 
$f_{ac} = (\gamma /2\pi) B\sqrt{1+(\mu_{\rm 0} M_\text{s}/2B_{\text{ex}})}$ and 
$f_{op} = (\gamma/2\pi)\sqrt{2B_{\text{ex}}\mu_{\rm 0} M_{\text{s}}\left(1-\left(B/2B_{\text{ex}}\right)^{2}\right)}$ where $\gamma$, $M_\text{s}$, $\mu_{\rm 0}$ and $B_{\text{ex}}$ are the gyromagnetic ratio, saturation magnetisation, free space permeability and interlayer exchange field respectively - we notice that there is a very subtle non-linear component for the acoustic mode for low frequency region which we cannot account for by our macrospin model. In Fig. 1(d), we present our numerical solutions of our eigenvalue problem (see details in Ref. [18]) to show good agreement between experimental observation in our device and model calculations. These demonstrations warrant that we are able to excite and measure both acoustic and optical modes in our SyAF STT-FMR devices with dual spin excitation layers. 

Individual curves are further analysed by the following Lorentzian functions to decompose their symmetric ($V_\text{sym}$) and anti-symmetric ($V_\text{asy}$) components.
\begin{equation}
        V_\text{dc} = V_\text{sym}\frac{\Delta B^2}{(B - B_\text{res})^2 + \Delta B^2} + V_\text{asy}\frac{\Delta B(B -  B_\text{res})}{(B - B_\text{res})^2 + \Delta B^2} 
    \end{equation}
Here, $B_\text{res}$ and $\Delta B$ are resonance field and half width at half maximum linewidth of resonance respectively. Typical FMR curve fit results are shown in Fig. 2(a) which represent excellent fit quality that is also the case for other curve fit analysis throughout this study. Figure 2(b) displays angular dependence of both $V_\text{sym}$ and $V_\text{asy}$ measured for acoustic modes while changing $\phi$. Both components clearly show sin2$\phi$sin$\phi$ angular dependence which can be explained by parallel pumping as follows. The angular dependence of observed rectification voltages can be interpreted by the product of the AMR angular dependence (sin2$\phi$) and torque symmetry\cite{MeckingPRB2007,FangNNano2011} (also see SM). We therefore divide both $V_\text{sym}$ and $V_\text{asy}$ by sin2$\phi$ to reveal the torque symmetry, which is shown as Fig. 2(c) that strongly suggests the torque symmetry for the acoustic mode excitation being the form of sin$\phi$. This is the case of parallel pumping, i.e. the torque is largest when applied magnetic field and oscillating excitation field are colinear ($\phi=90^\circ$ in our case since both $\boldsymbol{B_\text{1,2}}$ and $\boldsymbol{s_\text{1,2}}$ are along this direction). This is remarkably different from the perpendicular pumping nature of acoustic modes in SyAFs which gives cos$\phi$ dependence when they are homogeneously excited as discussed earlier. To determine the efficiency of spin-charge conversion in our devices, we quantified the spin-Hall angle ($\theta_{\text{SH}}$) by using the following expression of $V_\text{sym}$ for the acoustic modes; this is obtained by solving the LLG equation with coupled moments excited by opposite spin torques at the macrospin limit as shown in SM; 
\[ \begin{array}{c}\tag{4}
 V_\text{sym} =\ddfrac{-\Delta R_{\text{AMR}}C_0B_\text{SHE}\cos2\phi_\text{c}\sin\phi_\text{c}}{8\Delta B\sqrt{1+(\mu_{\rm 0} M _{\rm s}/2B_{\rm ex})} I_0 R_{\text{sample}}} P_\text{input}\sin2\phi\sin\phi
\end{array}
\] 
where $B_{\text{SHE}}$ is the field due to SHE and is defined as $\hbar\eta_{\text{Ta}}\theta_{\text{SH}}I_0/2eM_\text{s}wd_{\text{FM}}d_{\text{Ta}}$. Here, $\hbar$, $\eta_{\text{Ta}}$, $I_0$, $e$, $w$, $d_{\text{FM}}$, $d_{\text{Ta}}$, $\Delta R_{\text{AMR}}$, $\phi_{\text{c}}$, $P_{\rm input}$, $R_{\rm sample}$ and $C_0$ refer to the reduced Planck's constant, shunt ratio of current in Ta layer, current amplitude in the device, elementary charge, width of microbar, the thickness of the NiFe and Ta layers, AMR resistance change, cant angle, microwave power at the source, device resistance and the microwave calibration factor to convert the microwave power at the source to that in the device (see SM) respectively. This equation suggests linear relationship between $V_\text{sym}$ and $P_\text{input}$, which can be observed in Fig. 2(d). Using the slope fit by the $V_\text{sym}$ plot and Eq. (4), we extracted the magnitude of $\theta_{\text{SH}}$ of our Ta layers to be 0.1, which is consistent with previous studies\cite{LiuScience2012,SagastaPRB2018}. This can strongly suggest that the source of generating $ V_\text{sym}$ in our measurements is due to SHE in the Ta layers and we therefore conclude that the parallel pumping of SyAF acoustic modes can be achieved by dual spin sources with opposite spin polarisation. 

We expand our analysis to the optical modes in our devices, which is summarised in Fig. 3. The rectification voltages generated by optical modes have the following forms (see full derivations in SM).
\[ \begin{array}{lcl}\tag{5}
 V_\text{sym} = -\ddfrac{\Delta R_{\rm AMR}}{8  \Delta B}\sqrt{\ddfrac{2B_{\rm ex}}{\mu_{\rm 0} M_{\rm s}}} I_0B_{\rm SHE}\sin2\phi_{\rm c}\sin\phi_{\rm c}\cos 2\phi\cos \phi;\\
V_\text{asy} = \ddfrac{\Delta R_{\rm AMR}}{8  \Delta B} I_0B_{\rm Oe}  \sin2\phi _{\rm c}\cos 2\phi \cos \phi
 \end{array}
\] 
Here, $ B_{\text{Oe}}=\mu_{\text{0}}\eta_{\text{asy}}I_0/2w$ is the Oersted field due to current flowing in Ta layer with the parameter $\eta_{\text{asy}}$ being the asymmetry factor of electric currents between the Ta and NiFe layers. AMR symmetry for the optical mode is given by $\cos2\phi$ in these cases and therefore we divide experimentally-observed $V_\text{sym}$ and $V_\text{asy}$ by this to reveal the excitation torque symmetry which is shown in Fig. 3(b). Unlike the torque symmetry for the acoustic mode (Fig. 2(c)), now we confirm the symmetry for the optical mode is mainly described by cos$\phi$, which in our case indicates the perpendicular nature of spin-wave excitations, i.e. the torque (hence spin-wave excitation efficiency) is maximised when the oscillating fields and dc magnetic field are perpendicular to each other. We note that there might be some higher-order terms in this angular dependence that our model cannot capture. However, we emphasise that the main angular dependence in our experiments is clearly reproduced by our model, for both acoustic and optical mode excitations. The sign flip between $V_\text{sym}$ and $V_\text{asy}$ for the optical mode, which is predicted by our macrospin model i.e. in Eq. (5), is also clearly demonstrated in our STT-FMR experiments. $\theta_{\text{SH}}$ extracted using the slope of our experimental data in Fig. 3(c) is 0.1, showing good agreement with one extracted by using the acoustic mode resonances. 

Finally, we discuss the frequency dependence of the torque symmetry for both modes. In order to quantitatively discuss this, we fit the torque symmetry data (e.g. Fig. 2(c)) by $A$sin$\phi$+$B$cos$\phi$ to capture both parallel and perpendicular natures of spin-wave excitations. Using the prefactors $A$ and $B$, we define the angle $\Theta$ = arctan($A/B$) which indicates the degree to which spin-waves are excited by parallel or perpendicular pumping configuration; $\Theta$ close to $\pm 90 ^\circ$ (0$^\circ$) suggests the parallel (perpendicular) pumping nature in this definition. We have extracted $\Theta$ from our angular dependent measurements and show them in Figs 4(a) and (b) for the acoustic and optical modes respectively. We observe consistent behaviours of the torque symmetry for both modes as a function of frequency. This confirms the robust parallel (perpendicular) pumping nature of acoustic (optical) modes excited in our SyAF devices with dual spin torque sources using our parity control. 

In summary, we demonstrate in this study that local spin torque excitations can convert mode excitation symmetry between perpendicular and parallel pumping configurations in SyAF STT-FMR devices. We show this by using two spin-wave modes, acoustic and optical, both clearly exhibiting the torque symmetry change from the ones expected in spin-wave excitations with homogeneous fields. We also present full expressions of rectification voltages in SyAF STT-FMR devices with dual spin excitations, which supports our experimental observation as well as allows parameter extractions such as the spin-Hall angle using the rectification voltages. We envisage that the control of spin-wave excitations in STT nano-devices will be useful for future spintronic and magnonic nano-devices.     

We thank Kei Yamamoto for fruitful discussions on this topic. A. S. thank EPSRC for their supports through NPIF EPSRC Doctoral studentship (EP/R512400/1). This project was supported in part by CSRN, CSIS and UCL-Tohoku Strategic Partner Funds.

\section*{Data Availability}
The data that support the findings of this study are available from the corresponding author upon reasonable request.

\newpage
\begin{figure}[h]
\begin{center}
\includegraphics[scale=1.2]{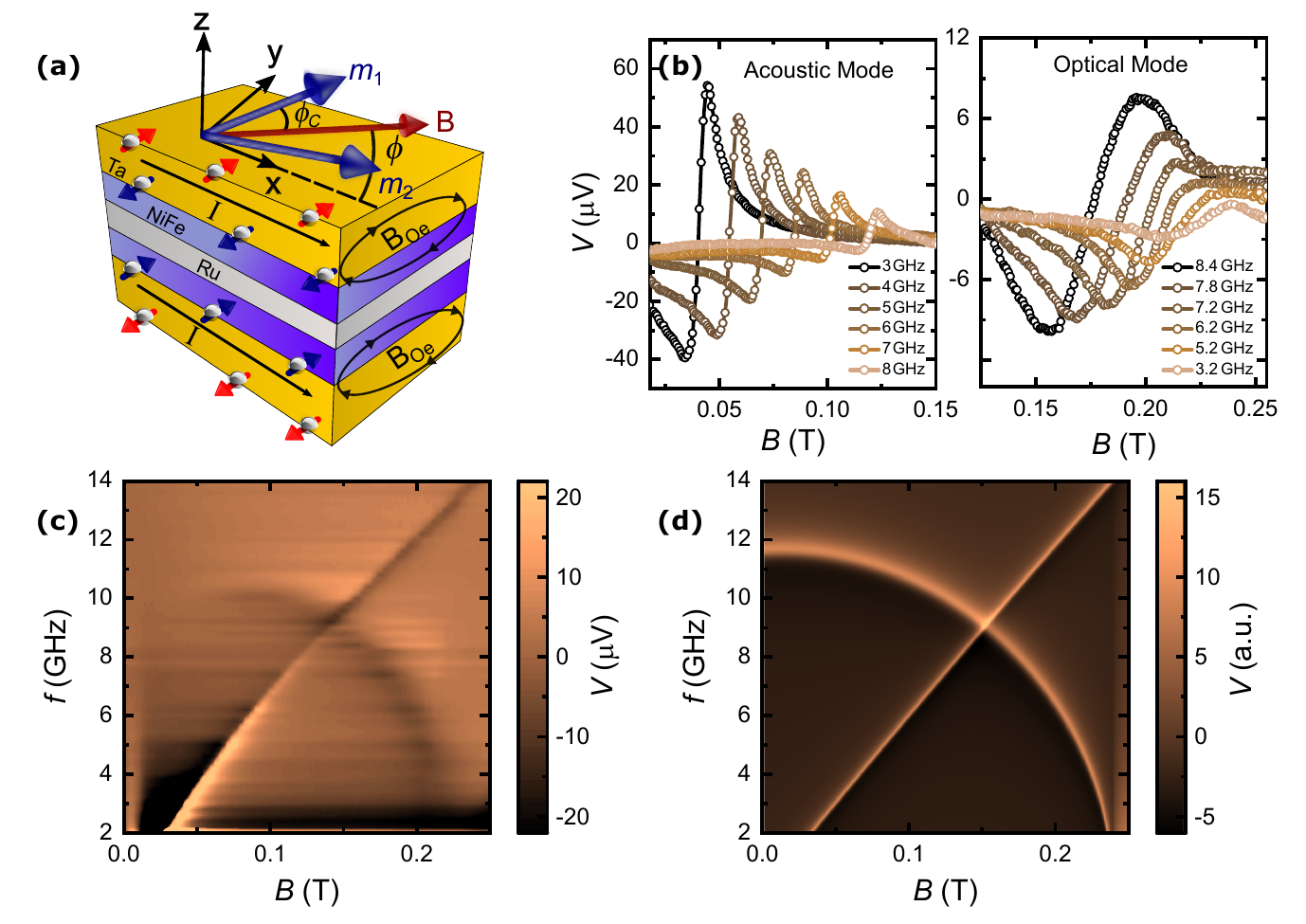}
\end{center}
\caption{(a) A schematic of the sample geometry used for STT-FMR measurements in our study. (b) $V$ obtained at different frequencies for acoustic and optical modes in our device for $\phi = 55^\circ$ (the values for some frequencies have been scaled to show them properly). (c) A 2D colorplot of $V$ as a function of applied field and frequency, measured for $\phi = 55^\circ $. (d) Numerical results calculated for the experimental conditions as in (d).}
\end{figure}

\newpage
\begin{figure}[h]
\begin{center}
\includegraphics[scale=0.15]{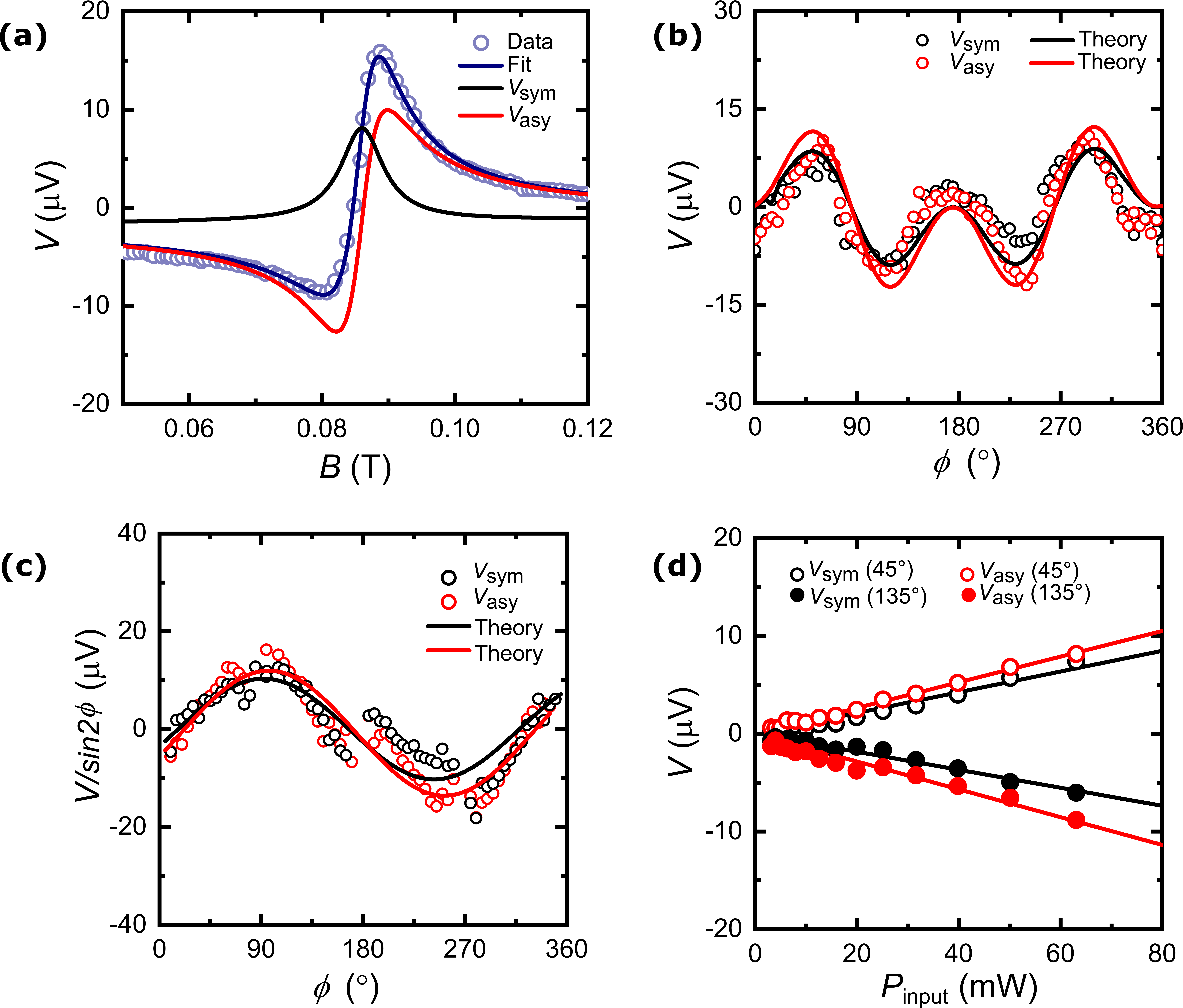}
\end{center}
\caption{(a) FMR spectra measured for $f$ = 6 GHz and $\phi = 45^\circ $, together with fitting curves produced by using Eq. (3). (b) Angular dependence of $V_\text{sym}$ and $V_\text{asy}$ for the acoustic mode measured at 8 GHz. We also add best fit solid curves using our model. (c) The symmetry of torques obtained by dividing the Voltage by $\sin2\phi$. The dominant $\sin\phi$ dependence of torques confirms the parallel pumping configuration. 
(d) $V_\text{sym}$ and $V_\text{asy}$ as a function of $P$ for $\phi = 45^\circ$ and $135^\circ $ for 8 GHz . The solid lines are linear fit to the data.}
\end{figure}

\newpage
\begin{figure}[h]
\begin{center}
\includegraphics[scale=0.3]{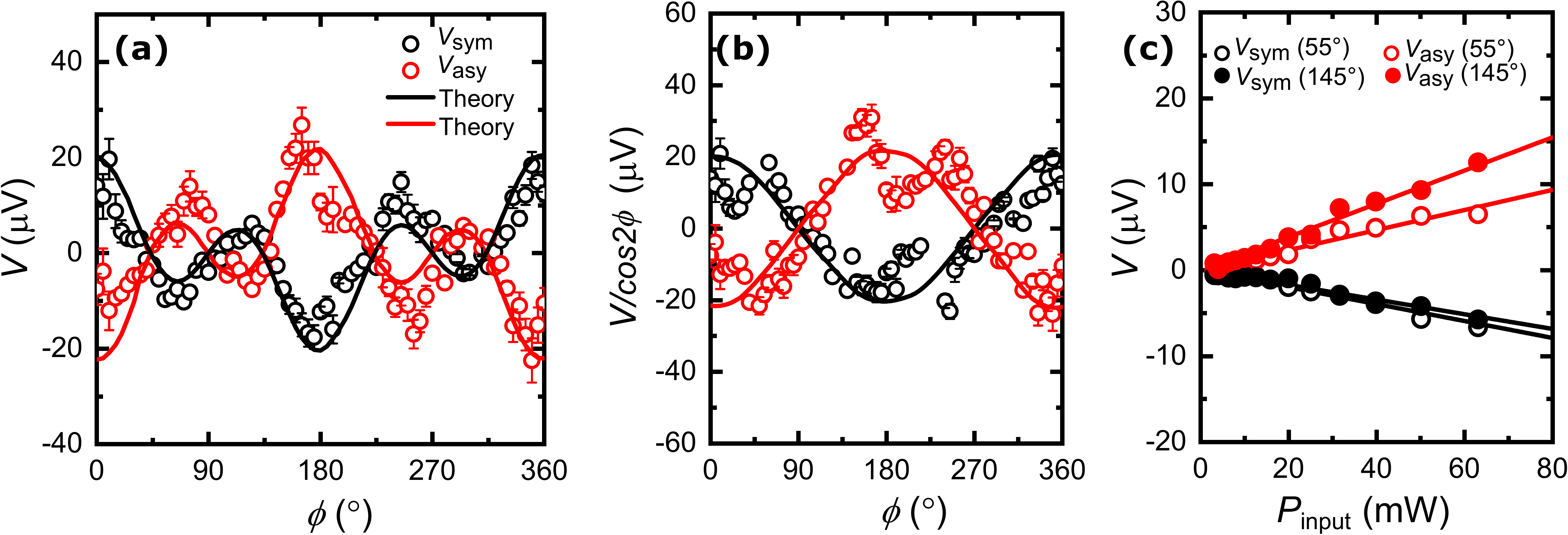}
\end{center}
\caption{(a)  Angular dependence of $V_\text{sym}$ and $V_\text{asy}$ for the optical mode measured at 8.5 GHz. The solid curves are obtained using Eq. (5). (b) The symmetry of torques obtained by dividing the Voltage by $\cos2\phi$. The dominant $\cos\phi$ dependence of torques confirms the perpendicular pumping configuration.  (c) The power dependence of $V_\text{sym}$ and $V_\text{asy}$ in the optical mode measured at $\phi = 55^\circ $and $145^\circ $ for $f$ = 8.5 GHz, together with linear fit results.}
\end{figure}

\newpage
\begin{figure}[h]
\begin{center}
\includegraphics[scale=0.33]{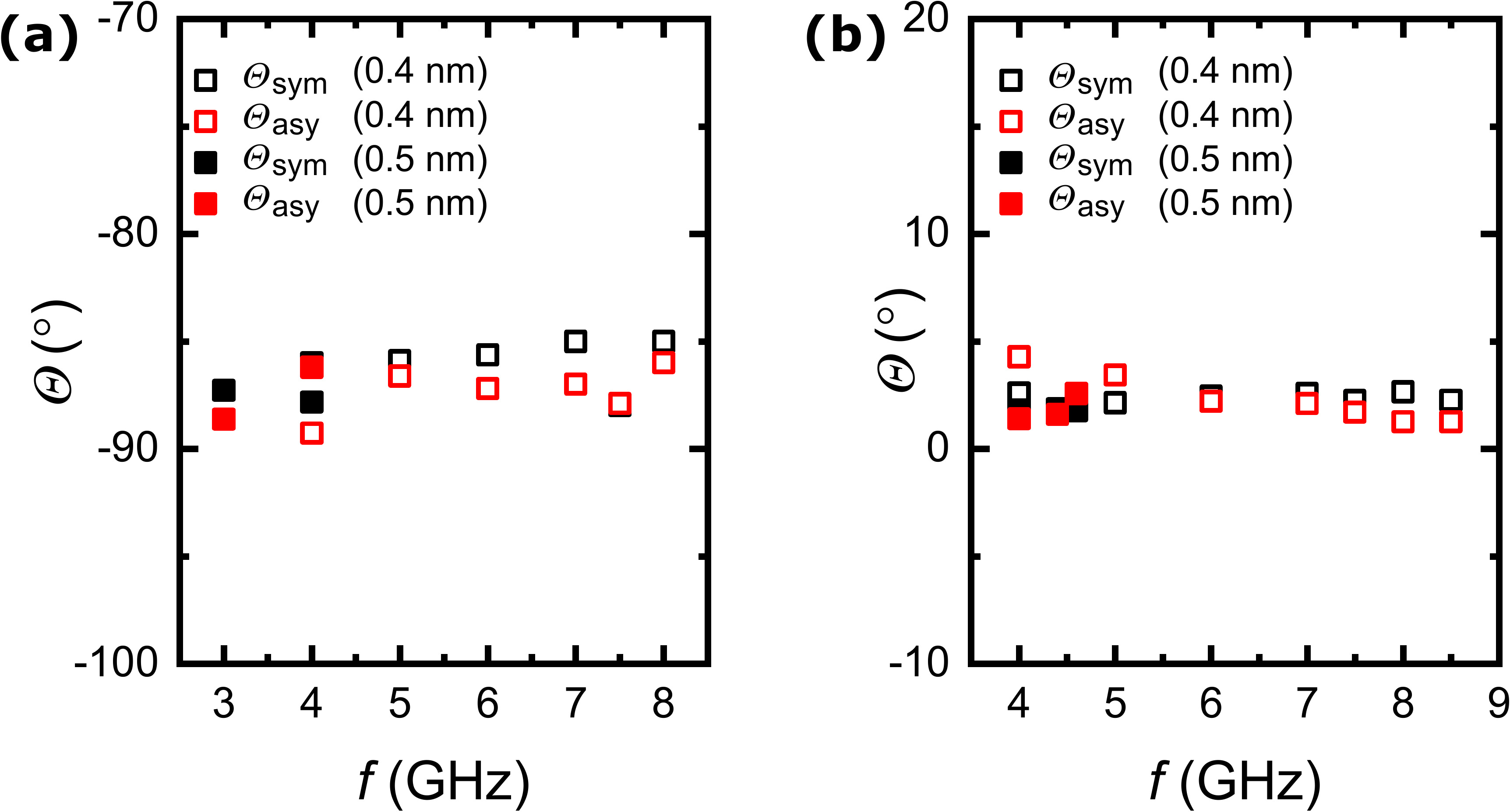}
\end{center}
\caption{$\Theta$ extracted as a function of frequency for (a) acoustic and (b) optical modes for different frequencies, in two samples with different Ru thicknesses (0.5 or 0.4 nm). $\Theta_\text{sym(asym)}$ represents the parameter extracted from the angular dependence of $V_\text{sym}$ ($V_\text{asy}$) for each frequency. These figures confirm that the acoustic and optical modes are excited by the torque symmetry of parallel and perpendicular pumping configurations respectively.}
\end{figure}

\pagebreak
\widetext
\begin{center}
\textbf{\large Supplementary Material for "Parity-controlled spin-wave excitations in synthetic antiferromagnets"}
\end{center}

\renewcommand{\theequation}{S\arabic{equation}}
\renewcommand{\thefigure}{S\arabic{figure}}
\setcounter{equation}{0}
\setcounter{figure}{0}
\renewcommand{\bibnumfmt}[1]{[S#1]}
\renewcommand{\citenumfont}[1]{S#1}

\def\ve{\varepsilon}
\def\unit#1{\ \mathrm{#1}}
\font\myfont=cmr12 at 8pt
\section{Parity and symmetry analysis of spin-wave modes in synthetic antiferromagnets}
To discuss the parity of spin-wave modes in synthetic antiferromagnets (SyAFs), we start with Landau-Lifshitz-Gilbert equations with coupled magnetic moments ($\boldsymbol{m_1}$ and $\boldsymbol{m_2}$) by the antiferromagnetic exchange interaction with the strength of $B_\text{ex}$:   
\begin{align}
\frac{d\boldsymbol{m_1}}{dt}=- \gamma \boldsymbol{m_1} \times (\boldsymbol{B}-B_\text{ex}\boldsymbol{m_2}-M_\text{s}(\boldsymbol{m_1}\cdot\boldsymbol{e_\text{z}})\boldsymbol{e_\text{z}})+\boldsymbol{\tau_1}
\end{align}
\begin{align}
\frac{d\boldsymbol{m_2}}{dt}=- \gamma \boldsymbol{m_1} \times (\boldsymbol{B}-B_\text{ex}\boldsymbol{m_1}-M_\text{s}(\boldsymbol{m_2}\cdot\boldsymbol{e_\text{z}})\boldsymbol{e_\text{z}})+\boldsymbol{\tau_2}.
\end{align}
Here, $t$, $\gamma$, $\boldsymbol{B}$, $M_\text{s}$, $\boldsymbol{e_\text{z}}$ and $\boldsymbol{\tau_{1(2)}}$ are time, the gyromagnetic ratio, the external magnetic field vector, the saturation magnetisation, the unit vector along the film growth direction and excitation torques for $\boldsymbol{m_1}$($\boldsymbol{m_2}$) respectively. For the sake of simplicity, we ignore the damping term to discuss the symmetry and parity of spin-wave modes in this section. We now substitute the static and dynamic components of $\boldsymbol{m_1}$ and $\boldsymbol{m_2}$ as  $\boldsymbol{m_1}=\boldsymbol{m_1^0}+\delta \boldsymbol{m_1} e^{i\omega t}$ 
and $\boldsymbol{m_2}=\boldsymbol{m_2^0}+\delta \boldsymbol{m_2} e^{i\omega t}$ respectively. Substituting these into Eqs. S1 and S2 and focusing on the first-order terms of $e^{i\omega t}$, we have:
\begin{align}
i\omega \delta \boldsymbol{m_1}=- \gamma \boldsymbol{m_1^0} \times (\boldsymbol{B_\text{eq}}\delta \boldsymbol{m_1}-B_\text{ex}\delta\boldsymbol{m_2}-M_\text{s}(\delta\boldsymbol{m_1}\cdot\boldsymbol{e_\text{z}})\boldsymbol{e_\text{z}})+\boldsymbol{\tau_1}
\end{align}
\begin{align}
i\omega \delta \boldsymbol{m_2}=- \gamma \boldsymbol{m_2^0} \times (\boldsymbol{B_\text{eq}}\delta \boldsymbol{m_2}-B_\text{ex}\delta\boldsymbol{m_1}-M_\text{s}(\delta\boldsymbol{m_2}\cdot\boldsymbol{e_\text{z}})\boldsymbol{e_\text{z}})+\boldsymbol{\tau_2}
\end{align}
Here, $\boldsymbol{B_{\text{eq}}}$ for $\boldsymbol{m_1}$($\boldsymbol{m_2}$) is represented by $\boldsymbol{B} - B_\text{ex}\cdot\boldsymbol{m_{2(1)}^0}$. Using the parity of optical and acoustic modes where the former (latter) is even for $\pi$ rotation (i.e. $C_2$ operation) about $\boldsymbol{B}$, we can select $\delta \boldsymbol{m_\text{op}}=\delta \boldsymbol{m_1}+C_2\delta \boldsymbol{m_2}$ and $\delta \boldsymbol{m_\text{ac}}=\delta \boldsymbol{m_1}-C_2\delta \boldsymbol{m_2}$ as new bases of the above coupled LLG equations to describe the optical and acoustic modes, respectively\cite{McNeillPRL2019}. These modes are excited by the corresponding torque terms which can be obtained also by the linear combination of original torques acting on $\boldsymbol{m_1}$ and $\boldsymbol{m_2}$. These linearly-combined torques are given by $\boldsymbol{\tau_\text{op}}=\boldsymbol{\tau_1}+C_2\boldsymbol{\tau_2}$ and $\boldsymbol{\tau_\text{ac}}=\boldsymbol{\tau_1}-C_2\boldsymbol{\tau_2}$ respectively. We now examine these torques to associate with our experimental observation. The individual torques can be written as $\boldsymbol{\tau_\text{i}}=\boldsymbol{m_\text{i}^0}\times \boldsymbol{B_\text{i}} +  \boldsymbol{m_\text{i}^0}\times (\boldsymbol{m_\text{i}^0}\times \boldsymbol{s_\text{i}}$) (i = 1, 2) where the first and second terms are field-like and damping-like torques acting on each moment with $\boldsymbol{B_\text{i}}$ being the effective field and $\boldsymbol{s_\text{i}}$ being the spin polarisation of the spin-transfer torque. Using these we write:
\begin{align}
\boldsymbol{\tau_\text{op}}=\boldsymbol{m_\text{1}^0}\times \boldsymbol{B_\text{1}} + \boldsymbol{m_\text{1}^0}\times(\boldsymbol{m_\text{1}^0}\times \boldsymbol{s_\text{1}})+C_2(\boldsymbol{m_\text{2}^0}\times \boldsymbol{B_\text{2}} + \boldsymbol{m_\text{2}^0}\times(\boldsymbol{m_\text{2}^0}\times \boldsymbol{s_\text{2}}));
\end{align}

\begin{align}
\boldsymbol{\tau_\text{ac}}=\boldsymbol{m_\text{1}^0}\times \boldsymbol{B_\text{1}} + \boldsymbol{m_\text{1}^0}\times(\boldsymbol{m_\text{1}^0}\times \boldsymbol{s_\text{1}})-C_2(\boldsymbol{m_\text{2}^0}\times \boldsymbol{B_\text{2}} + \boldsymbol{m_\text{2}^0}\times(\boldsymbol{m_\text{2}^0}\times \boldsymbol{s_\text{2}})).
\end{align}
Anti-ferromagnets with two identical moments has the even parity under $C_2$ operation for the static regime, i.e. $\boldsymbol{m_1^0}=C_2 \boldsymbol{m_2^0}$. Using this property, Eq. (S5) can be re-written as:
\begin{align}
\boldsymbol{\tau_\text{op}}=\boldsymbol{m_\text{1}^0}\times (\boldsymbol{B_\text{1}} + C_2\boldsymbol{B_\text{2}}) + \boldsymbol{m_\text{1}^0}\times \{\boldsymbol{m_\text{1}^0} \times (\boldsymbol{s_\text{1}}+C_2\boldsymbol{s_\text{2}})\}.
\end{align}
Likewise, we can obtain the torque expression for exciting the acoustic mode as:
\begin{align}
\boldsymbol{\tau_\text{ac}}=\boldsymbol{m_\text{1}^0}\times (\boldsymbol{B_\text{1}} - C_2\boldsymbol{B_\text{2}}) + \boldsymbol{m_\text{1}^0}\times \{\boldsymbol{m_\text{1}^0} \times (\boldsymbol{s_\text{1}}-C_2\boldsymbol{s_\text{2}})\}.
\end{align}
These two expressions represent the torque symmetry of each mode excitation against the external magnetic field direction which is the axis of the $C_2$ rotation. When both moments are excited by uniform spin excitation, namely with the condition of $\boldsymbol{B_\text{1}}=\boldsymbol{B_\text{2}}$ and $\boldsymbol{s_\text{1}}=\boldsymbol{s_\text{2}}$, we can arrive at the following conclusions.
(i) When $\boldsymbol{B_\text{1}}$ and $\boldsymbol{s_\text{1}}$ are symmetric for the $C_2$ rotation, $\boldsymbol{\tau_\text{op}}$ is maximised ($\boldsymbol{B_\text{1}} + C_2\boldsymbol{B_\text{2}}=2\boldsymbol{B_\text{1}}$) and $\boldsymbol{\tau_\text{ac}}=0$, and (ii) when When $\boldsymbol{B_\text{1}}$ and $\boldsymbol{s_\text{1}}$ are anti-symmetric for the $C_2$ rotation, $\boldsymbol{\tau_\text{op}} =0$ and $\boldsymbol{\tau_\text{ac}}$ is largest. Here the meaning of $\boldsymbol{B_\text{1}}$ and $\boldsymbol{s_\text{1}}$ being symmetric (anti-symmetric) for the $C_2$ rotation is the condition of $\boldsymbol{B_\text{1}} \| \boldsymbol{B}$ ($\boldsymbol{B_\text{1}} \perp \boldsymbol{B}$) which is in general termed as parallel (perpendicular) pumping configuration in magnetic dynamics. Altogether we can summarise that under a uniform excitation condition, the optical (acoustic) mode can be excited by parallel (perpendicular) pumping configuration. In the present study, we take one step further to control the excitation symmetry by designing the local spin excitation configuration. When $\boldsymbol{m_1}$ and $\boldsymbol{m_2}$ experience non-uniform spin excitations, here we show that the perpendicular/parallel pumping nature can be tuned in our experiments. On the extreme case where $\boldsymbol{B_\text{1}}=-\boldsymbol{B_\text{2}}$ and $\boldsymbol{s_\text{1}}=-\boldsymbol{s_\text{2}}$, we can predict that the optical (acoustic) mode is excited by perpendicular (parallel) pumping configuration. This is because the odd-even parity of the spin excitation is changed, leading to the parity change of excited modes by the spin torques with the spatial symmetry. 
\section{Rectification voltages arising from acoustic and optical modes in synthetic antiferromagnets}
Using the Kittel and Neel vector definitions (${\bf m}$ = (${\bf m}_1 + {\bf m}_2$)/2 and ${\bf n}$ = (${\bf m}_1 - {\bf m}_2$)/2), we rewrite the LLG equations as
\begin{align}
&\frac{d{\bf m}}{dt} = -\Omega _{\rm L} {\bf m}\times \boldsymbol{u} + \Omega _{\rm B} \left\{ \left[ {\bf m} \times ( {\bf m}\cdot {\bf e}_{\rm z}) {\bf e}_{\rm z} \right] + \left[ {\bf n} \times ( {\bf n}\cdot {\bf e}_{\rm z}) {\bf e}_{\rm z} \right] \right\} \notag \\
&\hspace{5cm}+\alpha \left[ {\bf m}\times \frac{d{\bf m}}{dt} + {\bf n}\times \frac{d{\bf n}}{dt} \right], \label{eq:dmdt}   \\
&\frac{d{\bf n}}{dt} = -\Omega _{\rm L} {\bf n}\times \boldsymbol{u} + \Omega _{\rm B} \left\{ \left[ {\bf n} \times ( {\bf m}\cdot {\bf e}_{\rm z}) {\bf e}_{\rm z} \right] + \left[ {\bf m} \times ( {\bf n}\cdot {\bf e}_{\rm z}) {\bf e}_{\rm z} \right] \right\} \notag \\
&\hspace{5cm} + 2 \Omega _{\rm ex} \left( {\bf n}\times {\bf m} \right) +\alpha \left[ {\bf n}\times \frac{d{\bf m}}{dt} + {\bf m}\times \frac{d{\bf n}}{dt} \right]. \label{eq:dndt}
\end{align}
Here, $\Omega _{\rm L} = \gamma B$, $\Omega _{\rm B}=\gamma B_{\rm S}$ (where $B_{\rm S} = \mu_{\rm 0} M_{\rm s}$) and $\Omega _{\rm ex} = \gamma B_{\rm ex}$ respectively and $\boldsymbol{u}$ is the unit vector along the direction of $\boldsymbol{B}$ which in our case is applied along the in-plane $x$ direction ($\boldsymbol{u} = {\bf e}_{\rm x}$) as shown in Fig. S1 - ${\bf e}_{\rm i}$ is the unit vector expression for each axis $i$. Vector components of ${\bf m}$ and ${\bf n}$ can be linearized as the following equations:
\begin{align}
&m_{\rm x}(t) = m_{\rm x0} + \delta m_{\rm x}(t) +\cdots \notag \\
&\hspace{0.8cm}= \cos \phi_{\text{c}} + \delta m_{\rm x} (t), \label{eq:dmx} \\
&m_{\rm y}(t) = m_{\rm y0} + \delta m_{\rm y}(t) +\cdots \notag \\
&\hspace{0.8cm}= \delta m_{\rm y} (t), \label{eq:dmy} \\
&m_{\rm z}(t) = m_{\rm z0} + \delta m_{\rm z}(t) +\cdots \notag \\
&\hspace{0.8cm}= \delta m_{\rm z} (t), \label{eq:dmz} \\
&n_{\rm x}(t) = n_{\rm x0} + \delta n_{\rm x}(t) +\cdots \notag \\
&\hspace{0.8cm}= \delta n_{\rm x} (t), \label{eq:dnx} \\
&n_{\rm y}(t) = n_{\rm y0} + \delta n_{\rm y}(t) +\cdots \notag \\
&\hspace{0.8cm}= \sin \phi_{\text{c}} + \delta n_{\rm y} (t), \label{eq:dny} \\
&n_{\rm z}(t) = n_{\rm z0} + \delta n_{\rm z}(t) +\cdots \notag \\
&\hspace{0.8cm}= \delta n_{\rm z} (t). \label{eq:dnz} 
\end{align}
By using the linearization of Eqs. (\ref{eq:dmdt}) and (\ref{eq:dndt}) and solving the first order term, where ($\delta m_{\rm y}$, $\delta m_{\rm z}$, $\delta n_{\rm x}$) and ($\delta m_{\rm x}$, $\delta n_{\rm y}$, $\delta n_{\rm z}$) are coupled and representing the motion of  acoustic and optical modes respectively, we find the matrix form of two equations by using the Fourier transformation ($f(t)=\int ^{\infty }_{-\infty } d\Omega f(\Omega ) \exp (i\Omega t)/2\pi $) as,
\begin{align}
&\left( 
\begin{array}{ccc}
i\Omega & i\alpha \Omega \cos \phi_{\text{c}} + \Omega _{\rm B} \cos \phi_{\text{c}} +\Omega_{\rm L}& 0 \\
-i\alpha \Omega \cos \phi_{\text{c}}  - \Omega_{\rm L} & i\Omega & i\alpha \Omega \sin \phi_{\text{c}} \\
0 & -i\alpha \Omega  \sin \phi_{\text{c}} -(2\Omega_{\rm ex}+\Omega _B) \sin \phi_{\text{c}} & i\Omega 
\end{array}
\right) \left( 
\begin{array}{c}
\delta m_{\rm y} \\
\delta m_{\rm z} \\
\delta n_{\rm x} 
\end{array}
\right) = {\boldsymbol \tau}_{\rm m}, \label{eq:acmat} \\
&\left( 
\begin{array}{ccc}
i\Omega & 0 & -i\alpha \Omega \sin \phi_{\text{c}}  -\Omega _{\rm B} \sin \phi_{\text{c}}  \\
0 & i\Omega & i\alpha \Omega  \cos \phi_{\text{c}} -(2\Omega_{\rm ex}-\Omega _{\rm B})\cos \phi_{\text{c}} + \Omega_{\rm L} \\
i\alpha \Omega \sin \phi_{\text{c}}  +2\Omega_{\rm ex} \sin \phi_{\text{c}} & -i\alpha \Omega \cos \phi_{\text{c}} -\Omega_{\rm L} +2\Omega_{\rm ex} \cos\phi_{\text{c}} & i\Omega 
\end{array}
\right)  \notag \\
&\hspace{10cm} \left( 
\begin{array}{c}
\delta m_{\rm x} \\
\delta n_{\rm y} \\
\delta n_{\rm z} 
\end{array}
\right) = {\boldsymbol \tau}_{\rm n}. \label{eq:opmat}
\end{align}
The right hand side of the equations (${\bf \tau }_{\rm m}, {\bf \tau }_{\rm n}$) are magnetic torques on each coupled moment to excite magnetisation dynamics. The complex resonance frequency for both modes ($\tilde{\Omega }_{\rm ac}$ and $\tilde{\Omega }_{\rm op}$) can be obtained by using the determinant of the  3 $\times $ 3 matrix in the left hand side as,
\begin{align}
&\tilde{\Omega }_{\rm ac} = \sqrt{\Omega_{\rm L}^2(1+\frac{\Omega _{\rm B}}{2\Omega_{\rm ex}} )} + i\cdot \frac{1}{2}\alpha \left( \frac{\Omega_{\rm L}^2}{2\Omega_{\rm ex}} + \Omega _{\rm B} +2 \Omega_{\rm ex} \right)  , \label{eq:omega_ac} \\
&\tilde{\Omega }_{\rm op} =  \sqrt{\frac{\Omega _{\rm B}}{2\Omega_{\rm ex}}\left( 4\Omega_{\rm ex}^2 - \Omega_{\rm L}^2\right)}  +i\cdot \frac{1}{2}\alpha \left( \Omega _{\rm B} + \frac{1}{2\Omega_{\rm ex}}(4\Omega_{\rm ex}^2 -\Omega_{\rm L}^2 ) \right)   ,   \label{eq:omega_op}
\end{align}
Here, the relation $\cos \phi_{\text{c}} = \Omega_{\rm L} / 2\Omega_{\rm ex}$ is used. The real part of these equations represents the resonance field for each mode of SyAFs that we show in our main text. We note that these equations are consistent with earlier work by Chiba {\it et al.} without considering the spin-pumping effect \cite{ChibaPRB2015}.
\begin{figure}[h]
\begin{center}
\includegraphics[width=14cm,keepaspectratio,clip]{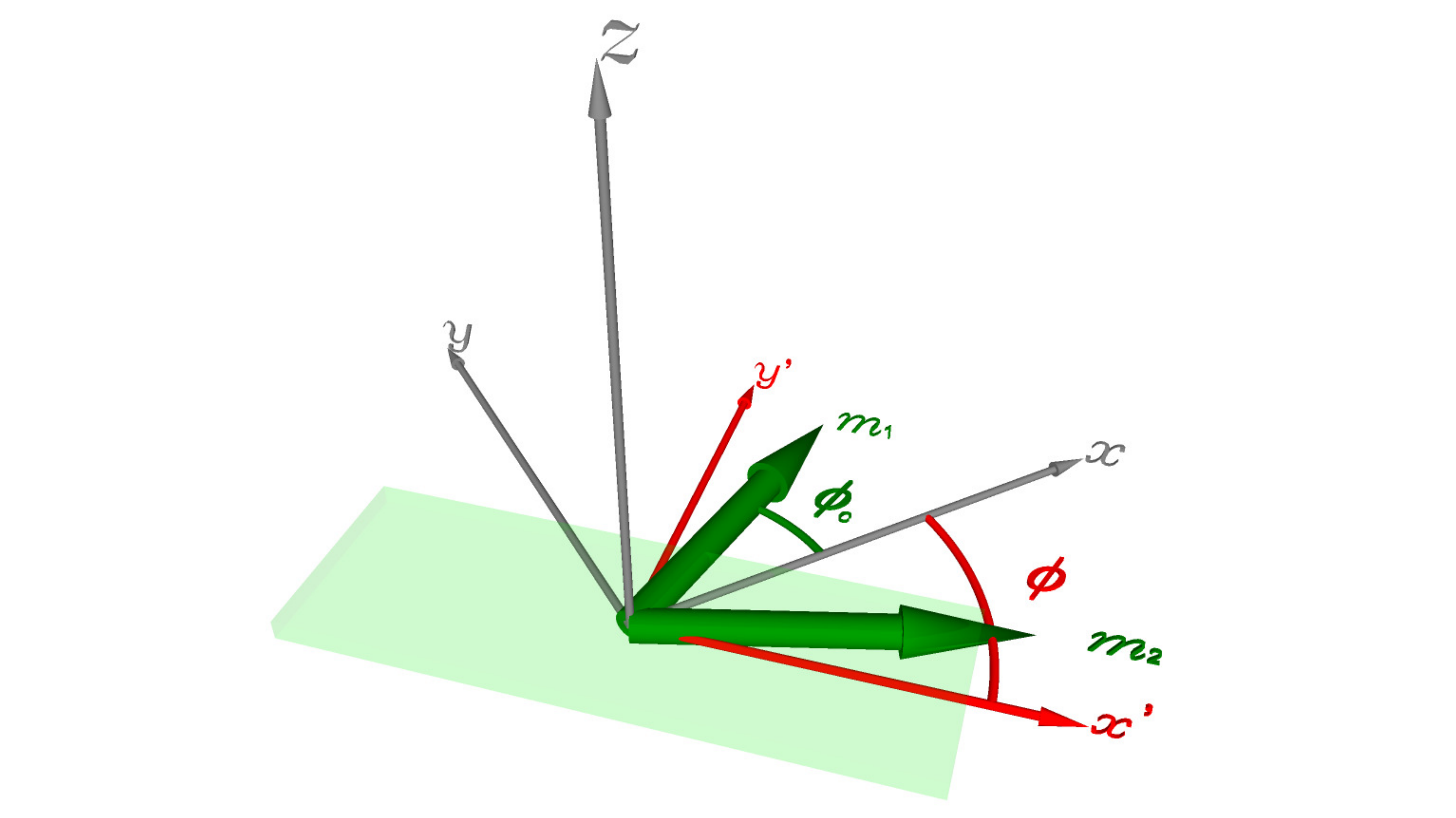}
\end{center}
\caption{A schematic of the coordinate system used in our macro-spin model. In this geometry, we have two Cartesian coordinate systems defined by the current flow ($x^{\prime }$) direction and the external magnetic field ($x$) direction. The angle between the two is defined by $\phi$ and the cant angle is defined by $\phi_{\text{c}} $. Note that the coordinates used are different from the ones in Fig. 1(a) in the main manuscript.}
\label{fig:coordinate}
\end{figure}
Now we consider current-induced torques and rectification voltages in our SyAF devices. We assume that an electric current $\boldsymbol{I}$ flows along the $x^{\prime }$ direction as defined in Fig. S1. $\boldsymbol{B}$ is applied along the $x$ direction from which we define $\phi$ with respect to the $x^{\prime }$ direction. In this coordinate, an Oersted field is generated along the $y^{\prime }$ direction, exerting torques (${\boldsymbol \tau}_{{\rm Oe}, 1(2)}$) on $\boldsymbol{m_1}$ and $\boldsymbol{m_2}$ as
\begin{align}
{\boldsymbol \tau}_{{\rm Oe}, 1(2)} &= \gamma B _{\rm Oe, 1(2)} {\bf m}_{1(2)} \times {\bf e}_{\rm y^{\prime }} \notag \\
&=\gamma B_{\rm Oe, 1(2)} \cos (\phi_{\text{c}} + \phi) {\bf e}_{\rm z}.
\end{align}
Here $B_{\rm Oe, 1(2)}$ can be approximated as $\mu_{\text{0}}\eta_{\text{asy}}I/2w$ with $\mu _0$ and $w$ being the permeability of free space and the width of the wire respectively. The parameter $\eta_{\text{asy}}$ can be obtained by estimating currents flowing above and below the magnetic layers. When the stacking structure is symmetric, we have $B_{\rm Oe, 1}=-B_{\rm Oe, 2}$. Therefore, the torques acting on ${\bf m}$ and ${\bf n}$ in Eqs. (\ref{eq:acmat}) and (\ref{eq:opmat}) are obtained as:
\begin{align}
&{\boldsymbol \tau}_{\rm m} = \gamma \left( 
\begin{array}{c}
0 \\
-B_{\rm Oe} \sin \phi_{\text{c}} \sin \phi \\
0 
\end{array}
\right) , \label{eq:tau_m} \\
&{\boldsymbol \tau}_{\rm n} = \gamma \left( 
\begin{array}{c}
0 \\
0 \\
B_{\rm Oe} \cos \phi_{\text{c}} \cos \phi  
\end{array}
\right) . \label{eq:tau_n}
\end{align} 
In addition to these torques, spin-transfer torques can be generated by the spin-Hall effect (SHE) from an adjacent nonmagnetic layer. Using our current direction which defines spin polarisation ${\boldsymbol \sigma }$ of the spin currents generated by the SHE as ${\boldsymbol \sigma }$ = ${\bf e}_{\rm y^{\prime }}$, we write the spin-transfer torques acting on two moments ($i=1,2$) as:
\begin{align}
{\boldsymbol \tau}_{{\rm SHE}, i} = \gamma B_{\rm SHE, i} \left[ {\bf m}_i \times ( {\bf e}_{\rm y^{\prime }} \times {\bf m}_i ) \right] ,
\end{align}
where $B_{\rm SHE, i}$ is an effective magnetic field proportional to both $I$ and the spin-Hall angle ($\theta_{\rm SH}$) with its magnitude $B_{\text{SHE}}$ given as 
$B_{\text{SHE}} = \hbar\eta_{\text{Ta}}\theta_{\text{SH}}I/2eM_\text{\rm s}wd_{\text{FM}}d_{\text{Ta}}$ where $\hbar$, $\eta_{\text{Ta}}$, $e$, $M_{\rm S}$, $d_{\text{FM}}$ and $d_{\text{Ta}}$ refer to the reduced Planck's constant, shunt ratio of current in Ta layer($I_{\rm Ta}/I$), elementary charge, saturation magnetisation and thickness of the NiFe and Ta layers respectively.

In the symmetric sample stack case, the relationship of $B_{\rm SHE, 1} = - B_{\rm SHE, 2}$ is established, giving the torque expressions for both vectors as 
\begin{align}
&{\boldsymbol \tau}_{\rm SHE, m} = \gamma \left( 
\begin{array}{c}
-B_{\rm SHE} \sin \phi_{\text{c}} \cos \phi_{\text{c}} \sin \phi \\
0 \\
B_{\rm SHE} \sin ^2 \phi_{\text{c}} \sin \phi 
\end{array}
\right) , \label{eq:tau_mSHE} \\
&{\boldsymbol \tau}_{\rm SHE, n} = \gamma \left( 
\begin{array}{c}
-B_{\rm SHE} \sin \phi_{\text{c}} \cos \phi_{\text{c}} \cos \phi \\
B_{\rm SHE} \cos ^2 \phi_{\text{c}} \cos \phi \\
0 
\end{array}
\right) . \label{eq:tau_nSHE}
\end{align} 
\newline
By using Eqs. (\ref{eq:acmat}), (\ref{eq:opmat}), (\ref{eq:tau_m}), (\ref{eq:tau_n}), (\ref{eq:tau_mSHE}), (\ref{eq:tau_nSHE}) the expression for $\delta m_{y}$, $\delta n_{\text{x}}$, $\delta m_{x}$ and $\delta n_{\text{y}}$ can be written as:
\begin{align}
&\delta m_{y}=\ddfrac{\gamma[(\Omega^2B_{\text{SHE}}\sin\phi_{\text{c}}\cos\phi_{\text{c}}\sin\phi) +i\cdot\Omega(\Omega_{\rm L}+\Omega_{B}\cos\phi_{\text{c}})(B_{\text{Oe}}\sin\phi_{\text{c}}\sin\phi)]}{-i\Omega(-\Omega_{\text{ac}}+i\cdot\delta_{\text{ac}}-\Omega)(\Omega_{\text{ac}}+i\cdot\delta_{\text{ac}}-\Omega)}\\
&\delta n_{\text{x}}= \ddfrac{\gamma[-\Omega_{\text{L}}(\Omega_{B}+2\Omega_{\rm ex})B_{\text{SHE}}\sin^2\phi_{\text{c}}\cos\phi_{\text{c}}\sin\phi-i \Omega\cdot B_{\text{Oe}}(\Omega_{\text{B}}+2\Omega_{\rm ex})\sin^2\phi_{\text{c}}\sin\phi]}{-i\Omega(-\Omega_{\text{ac}}+i\cdot\delta_{\text{ac}}-\Omega)(\Omega_{\text{ac}}+i\cdot\delta_{\text{ac}}-\Omega)}\\
&\delta m_{\text{x}} = \ddfrac{\gamma[\Omega^2(B_{\text{SHE}}\sin\phi_{\text{c}}\cos\phi_{\text{c}}\cos\phi+ i\Omega\cdot B_{\text{Oe}}\Omega_{\text{B}}\sin\phi_{\text{c}}\cos\phi_{\text{c}}\cos\phi)]}{-i\Omega(-\Omega_{\text{op}}+i\cdot \delta_{\text{op}}-\Omega)(\Omega_{\text{op}}+i\cdot\delta_{\text{op}}-\Omega)}\\
&\delta n_{\text{y}} = \ddfrac{\gamma[-\Omega^2B_{\text{SHE}}\cos^2\phi_{\text{c}}\cos\phi-i\Omega\cdot B_{\text{Oe}}\Omega_{\text{B}}\cos^2\phi_{\text{c}}\cos\phi]}{-i\Omega(-\Omega_{\text{op}}+i\cdot \delta_{\text{op}}-\Omega)(\Omega_{\text{op}}+i\cdot\delta_{\text{op}}-\Omega)}
\end{align}
Here, $\tilde{\Omega }_{\rm ac} = \pm \Omega _{\rm ac } + i\cdot \delta _{\rm ac}$, $\tilde{\Omega }_{\rm op} = \pm  \Omega _{\rm op } + i\cdot \delta _{\rm op}$ are used. The above equations can be simplified by calculating residue at two poles $(-\Omega_{\text{ac(op)}}+i\cdot \delta_{\text{ac(op)}})$ and $(\Omega_{\text{ac(op)}}+i\cdot\delta_{\text{ac(op)}})$ using the residue theorem given by:
\begin{align}
Res(f,c)= \lim_{z \to +c}(z-c) f(c).
\end{align}
Thus, we obtain the simplified form which can be expressed as below:
\begin{align}
&\delta m_{\rm y}(\Omega ) = \frac{\gamma }{\Omega _{\rm ac} +i\cdot \delta _{\rm ac} - \Omega }\cdot (a_{m_{\rm y}} + i \cdot b_{m_{\rm y}}) +\frac{\gamma }{-\Omega _{\rm ac} +i\cdot \delta _{\rm ac} - \Omega  }\cdot (-a_{m_{\rm y}} + i \cdot b_{m_{\rm y}}) , \notag \\
&\delta n_{\rm x}(\Omega ) = \frac{\gamma }{\Omega _{\rm ac} +i\cdot \delta _{\rm ac} - \Omega }\cdot (a_{n_{\rm x}} + i \cdot b_{n_{\rm x}}) +\frac{\gamma }{-\Omega _{\rm ac} +i\cdot \delta _{\rm ac} - \Omega }\cdot (-a_{n_{\rm x}} + i \cdot b_{n_{\rm x}}), \notag \\
&\delta m_{\rm x}(\Omega ) = \frac{\gamma }{\Omega _{\rm op} +i\cdot \delta _{\rm op} - \Omega}\cdot (a_{m_{\rm x}} + i \cdot b_{m_{\rm x}}) +\frac{\gamma }{-\Omega _{\rm op} +i\cdot \delta _{\rm op} - \Omega }\cdot (-a_{m_{\rm x}} + i \cdot b_{m_{\rm x}}), \notag \\
&\delta n_{\rm y}(\Omega ) = \frac{\gamma }{\Omega _{\rm ac} +i\cdot \delta _{\rm op} - \Omega }\cdot (a_{n_{\rm y}} + i \cdot b_{n_{\rm y}}) +\frac{\gamma }{-\Omega _{\rm ac} +i\cdot \delta _{\rm op} - \Omega }\cdot (-a_{n_{\rm y}} + i \cdot b_{n_{\rm y}}) . \notag
\end{align}
The components $a$ and $b$ are given by,
\begin{align}
&a_{m_{\rm y}} = \frac{1}{2}B_{\rm Oe} \frac{\Omega _{\rm ac}}{\Omega_{\rm L}} \sin \phi_{\text{c}} \sin \phi, \label{eq:amy}\\
&b_{m_{\rm y}} = -\frac{1}{2}B_{\rm SHE} \cos \phi_{\text{c}} \sin \phi_{\text{c}}\sin \phi ,\label{eq:bmy} \\
&a_{n_{\rm x}} = -\frac{1}{2\Omega _{\rm ac}}B_{\rm Oe} (\Omega _{\rm B} + 2\Omega_{\rm ex} ) \sin ^2 \phi_{\text{c}} \sin \phi, \label{eq:anx} \\
&b_{n_{\rm x}} = \frac{1}{2}B_{\rm SHE}\frac{\Omega_{\rm L}(\Omega _{\rm B}+2\Omega_{\rm ex})}{\Omega _{\rm ac}^2}   \sin ^2 \phi_{\text{c}} \cos \phi_{\text{c}}\sin \phi , \label{eq:bnx} \\
&a_{m_{\rm x}} = \frac{1}{2}B_{\rm Oe} \frac{\Omega _{\rm B}}{\Omega _{\rm op}}\sin \phi_{\text{c}} \cos \phi_{\text{c}}\cos \phi , \label{eq:amx} \\
&b_{m_{\rm x}} = -\frac{1}{2}B_{\rm SHE} \sin \phi_{\text{c}} \cos \phi_{\text{c}}\cos \phi  ,\label{eq:bmx}  \\
&a_{n_{\rm y}} = -\frac{1}{2}B_{\rm Oe}\frac{\Omega_{\text{B}}}{\Omega _{\rm op}} \cos^2 \phi_{\text{c}}\cos \phi   , \label{eq:any}
 \\
&b_{n_{\rm y}} = \frac{1}{2}B_{\rm SHE} \cos^2 \phi_{\text{c}} \cos \phi. \label{eq:bny}
\end{align}
Eqs. (\ref{eq:amy}) -- (\ref{eq:bnx}) are the amplitude for acoustic mode, which are proportional to $\sin \phi$. Eqs. (\ref{eq:amx}) -- (\ref{eq:bny}) are the amplitude for optical mode, which are proportional to $\cos \phi$. These essentially represent the torque symmetry we focus on our study.

The components $B_{\rm Oe}$ and $B_{\rm SHE}$ are both  proportional to an injected current $I$ into our devices, which has the time-varying form of $I(t) = I_0 \cos (\omega t)$. This time-varying current in our device produces magnetisation dynamics which causes another time-varying component in resistance due to AMR. As a result, there is a time-averaging component$< \Delta R (t) \cdot I(t) >$ when these two are mixed, which we measure in our homodyne detection approach \cite{TulapurkarNature2005,MeckingPRB2007,Fangnature2011}. A resistance change due to AMR has the following general form and we expand these for finding first-order time-varying components in both ${\bf m}$ and ${\bf n}$: 
\begin{align}
\Delta R (t) &= \frac{1}{4}\Delta R_{\rm AMR } \left[ ({\bf m}_1 \cdot {\bf e}_{\rm x^{\prime }})^2 + ({\bf m}_2 \cdot {\bf e}_{\rm x^{\prime }})^2 \right] \notag \\
&=\frac{1}{2}\Delta R_{\rm AMR} \left( \cos ^2 \phi (m_{\rm x}^2 + n_{\rm x}^2 ) + \sin ^2 \phi (m_{\rm y}^2 + n_{\rm y}^2 ) \right. \notag \\
&\hspace{2cm} \left.-2 \sin \phi \cos \phi (m_{\rm x}m_{\rm y} + n_{\rm x}n_{\rm y}) \right) \notag \\
&=\Delta R_{\rm AMR} \left( \cos ^2 \phi \cos \phi _{\rm c} \delta m_{\rm x}(t) \right. \notag \\
&\hspace{0.5cm} \left. + \sin ^2 \phi \sin \phi _{\rm c} \delta n_{\rm y} (t)- \sin \phi \cos \phi ( \cos \phi _{\rm c} \delta m_{\rm y}(t) + \sin \phi _{\rm c} \delta n_{\rm x}(t) \right) + \cdots . \label{eq:AMR}
\end{align} 
Equations (\ref{eq:amy}) to (\ref{eq:AMR}) allow to find the analytical expression of rectification voltages we measure in our SyAF devices. The anti-symmetric amplitude for the acoustic mode ($V_{\rm asy}^{\rm ac} $) can be derived as,
\begin{align}
V_{\rm asy}^{\rm ac} &= \frac{\gamma\Delta R_{\rm AMR}}{2\delta _{\rm ac} } I_0^2\cdot\frac{d}{dI} \left( a_{m_{\rm y}} \cos \phi _{\rm c} \sin \phi \cos \phi + a_{n_{\rm x}}\sin \phi _{\rm c} \sin \phi \cos \phi \right) \notag \\
&=-\frac{\gamma\Delta R_{\rm AMR}}{4 \Delta B_{\rm ac} (d\Omega _{\rm ac}/dB) } I_0^2\cdot \frac{dB_{\rm Oe}}{dI} \sin ^2 \phi \cos \phi \left( -\frac{\Omega _{\rm ac}}{\Omega_{\rm L}} \sin \phi _{\rm c} \cos \phi _{\rm c} + \frac{\Omega _{\rm B}+2\Omega_{\rm ex}}{\Omega _{\rm ac}}\sin ^3 \phi _{\rm c} \right) \notag \\ 
&= \frac{\Delta R_{\rm AMR}}{8  \Delta B_{\rm ac} } I_0B_{\rm Oe} \tan \phi _{\rm c} \cos 2\phi _{\rm c} \sin 2\phi \sin \phi .
\end{align}
Here, the linewidth $\delta _{\rm ac}$ is converted to  field-swept linewidth $\Delta B _{\rm ac}$ using the relation $\delta _{\rm ac} = (d\Omega _{\rm ac}/dB) \Delta B_{\rm ac}$, where $d\Omega _{\rm ac}/dB$ is given by $\gamma \sqrt{1+B _{\rm S}/(2B_{\rm ex})}$.
The relations $\cos \phi _{\rm c} = \Omega_{\rm L}/(2\Omega_{\rm ex})$, $\Omega _{\rm ac} = \Omega_{\rm L} \sqrt{1+\Omega _{\rm B}/(2\Omega_{\rm ex})}$ are used.  
Likewise, the symmetric amplitude for the acoustic mode $V_{\rm sym}^{\rm ac}$ is given as:
\begin{align}
V_{\rm sym}^{\rm ac} &= \frac{\gamma\Delta R_{\rm AMR}}{2\delta _{\rm ac} } I_0^2\cdot \frac{d}{dI}\left( b_{m_{\rm y}} \cos \phi _{\rm c} \sin \phi \cos \phi + b_{n_{\rm x}}\sin \phi _{\rm c} \sin \phi \cos \phi \right) \notag \\
&=-\frac{\gamma\Delta R_{\rm AMR}}{4 \Delta B_{\rm ac} (d\Omega _{\rm ac}/dB) } I_0^2\cdot \frac{dB_{\rm SHE}}{dI} \sin ^2 \phi \cos \phi \left( \sin \phi _{\rm c} \cos ^2 \phi _{\rm c} - \sin ^3 \phi _{\rm c} \right) \notag \\ 
&=-\frac{\Delta R_{\rm AMR}}{8  \Delta B_{\rm ac} \sqrt{1+B _{\rm S}/(2B_{\rm ex})}} I_0B_{\rm SHE} \sin \phi _{\rm c} \cos 2\phi _{\rm c} \sin 2\phi \sin \phi.
\end{align}
Using the relationships of $B _{\rm S}=\mu_0 M_\text{s}$ and $P_\text{input}=I_0^2R_\text{sample}/C_0$, we can find Eq. (4) in the main text. The symmetric and anti-symmetric amplitudes for the optical mode ($V_{\rm sym}^{\rm op}$ and $V_{\rm asy}^{\rm op}$) are:
\begin{align}
V_{\rm sym}^{\rm op} &= \frac{\gamma\Delta R_{\rm AMR}}{2\delta _{\rm op} } I_0^2\cdot \frac{d}{dI}\left( b_{m_{\rm x}} \cos\phi _{\rm c}  \cos^2 \phi + b_{n_{\rm y}}\sin\phi _{\rm c} \sin^2 \phi \right) \notag \\
&=-\frac{\gamma\Delta R_{\rm AMR}}{4 \Delta B_{\rm op} (d\Omega _{\rm op}/dB) } I_0^2\cdot \frac{dB_{\rm SHE}}{dI} \cos ^2 \phi_{\rm c} \sin \phi_{\rm c} \left( \cos ^3 \phi-\cos\phi \sin^2 \phi   \right) \notag \\ 
&=-\frac{\Delta R_{\rm AMR}}{8  \Delta B_{\rm op}(\Omega_{\rm B}\cdot\cos\phi_{\rm c}/\Omega_{\rm op}) } I_0^2\frac{dB_{\rm SHE}}{dI} \sin2\phi_{\rm c}\cos \phi _{\rm c} \cos 2\phi\cos \phi \notag \\
&=-\frac{\Delta R_{\rm AMR}}{8  \Delta B_{\rm op}}\sqrt{\frac{2B_{\rm ex}}{B_{\rm S}}} I_0B_{\rm SHE}\sin2\phi_{\rm c}\sin\phi_{\rm c}\cos 2\phi\cos \phi.  
\end{align}.
\begin{align}
V_{\rm asy}^{\rm op} &= \frac{\gamma\Delta R_{\rm AMR}}{2\delta _{\rm op} } I_0^2\cdot \frac{d}{dI}\left( a_{m_{\rm x}} \cos\phi _{\rm c}  \cos^2 \phi + a_{n_{\rm y}}\sin\phi _{\rm c} \sin^2 \phi \right) \notag \\
&=\frac{\gamma\Delta R_{\rm AMR}}{4 \Delta B_{\rm op} (d\Omega _{\rm op}/dB) } I_0^2\cdot \frac{dB_{\rm Oe}}{dI}\cdot\frac{\Omega_{\rm B}}{\Omega_{\rm op}} \cos ^2 \phi_{\rm c} \sin \phi_{\rm c} \left(  \cos^3 \phi -\cos\phi\sin ^2 \phi \right) \notag \\ 
&= \frac{\Delta R_{\rm AMR}}{8  \Delta B_{\rm op}(\Omega_{\rm B}\cdot\cos\phi_{\rm c}/\Omega_{\rm op})} I_0^2\frac{dB_{\rm Oe}}{dI}\cdot\frac{\Omega_{\rm B}}{\Omega_{\rm op}}  \sin2\phi _{\rm c} \cos \phi _{\rm c} \cos 2\phi \cos \phi \notag\\
&= \frac{\Delta R_{\rm AMR}}{8  \Delta B_{\rm op}} I_0B_{\rm Oe}  \sin2\phi _{\rm c}\cos 2\phi \cos \phi.
\end{align}
Here, we used the relation $\Omega_{\rm op} =\sqrt{\frac{\Omega_{\rm B}}{2\Omega_{\rm ex}}(4\Omega_{\rm ex}^2-\Omega_{\rm L}^2)}$ to convert  frequency swept linewidth $\delta_{\rm op}$ to field swept linewidth  $\Delta B _{\rm op}$.
\begin{figure}[h]
\centering
{\includegraphics[width=0.7\textwidth]{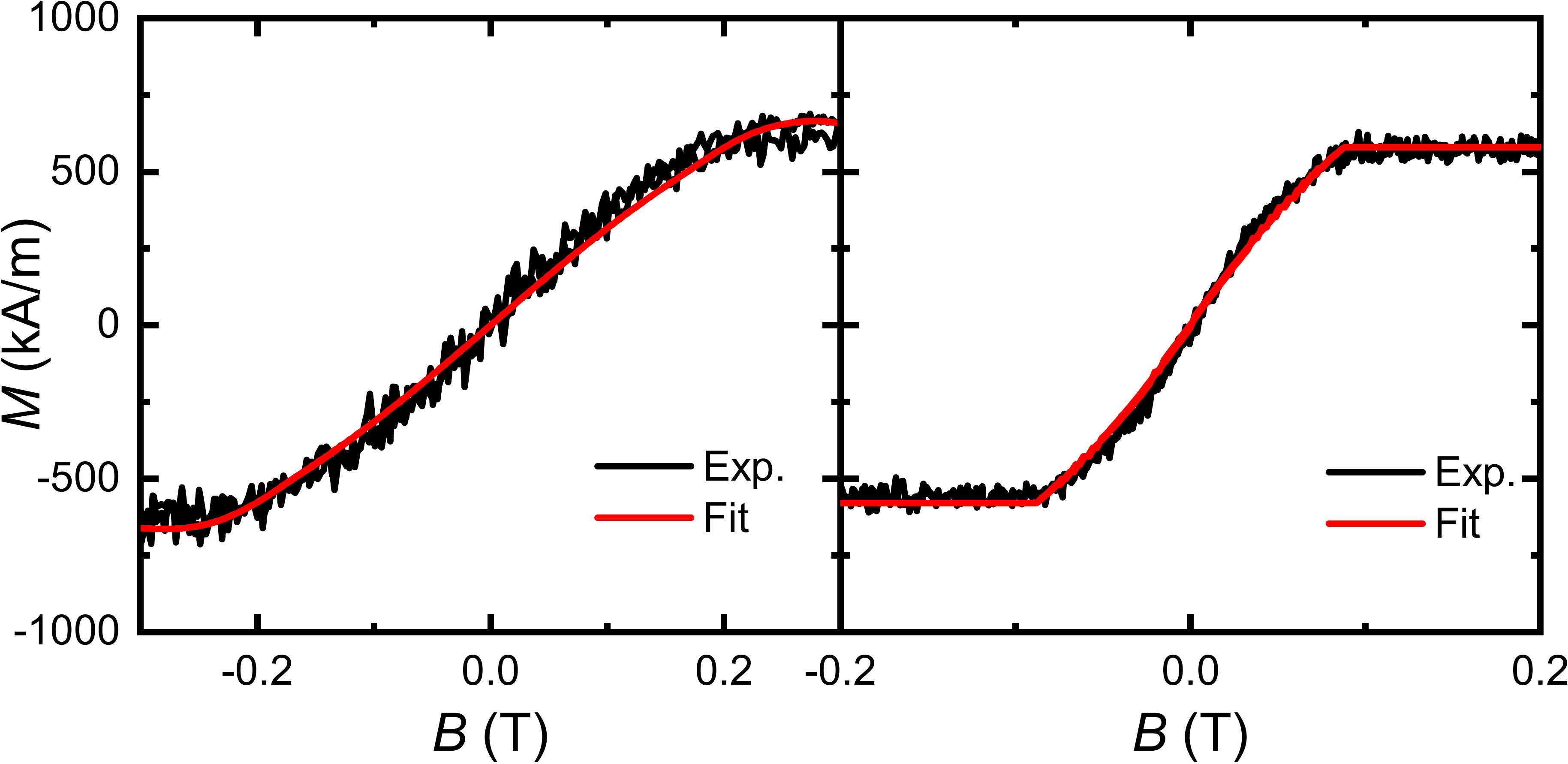}}
\caption{\label{fig:fig_2}(a-b) Magnetization curve of the Ta (5 nm)/ NiFe (5 nm)/Ru (t nm)/NiFe (5 nm)/Ta (5 nm) measured by vibrating sample magnetometer for (a) $t$ = 0.4 and (b) $t$ = 0.5. The black (red) curve is the experimental (calculation) results.}
\end{figure}
\section{Vibrating sample magnetometer characterisation and inter-layer exchange interaction}
Magnetic free energy of SyAFs can be described by the following equation\cite{biquadratic,martin,sorokin,PRB2020} including  bilinear and biquadratic exchange coupling contributions \cite{biquadratic}:
\begin{equation}
 F=\sum\limits_{j=1}^{2}\bigg[M_\text{s}\boldsymbol{B}\cdot\boldsymbol{m_j}+\frac{1}{2}M_\text{s}B_\text{s}\left(\boldsymbol{m_j}\boldsymbol{\cdot }\boldsymbol{z}\right)^2\bigg]+\frac{2J_\text{ex1}}{d}\boldsymbol{m_1}\cdot\boldsymbol{m_2}+\frac{2J_\text{ex2}}{d}\left(\boldsymbol{m_1}\cdot\boldsymbol{m_2}\right)^2 \label{eq:freeE}
\end{equation}
\noindent Here, the different terms $M_\text{s}$, ${\boldsymbol{m_{1(2)}}}$, ${\boldsymbol{B}}$, $B_\text{s}$, $J_\text{ex1(2)}$ are the saturation magnetisation,the unit vector of individual moments in a SyAF, external magnetic field vector,  demagnetisation field and  the linear and quadratic antiferromagnetic interlayer exchange coupling constants, respectively; the thickness of ferromagnet $d$ is identical for the present case. Figure~\ref{fig:fig_2} shows characterisation of two samples of Ru thickness 0.4 and 0.5nm  by a vibrating sample magnetometer. The red curves are obtained by minimising Eq. \ref{eq:freeE} iteratively to obtain $\phi(B)$ so as to obtain $M\left(B\right)$ given by  $M\left(B\right)=M_s\text{cos}\phi (B)$ \cite{sorokin,parkin,martin}. Here, $\phi(B)$ is the angle between the applied magnetic field direction and equilibrium direction of individual moments. The values of  the linear and quadratic exchange fields used to obtain the red curves are  120 (50)  $\pm$ 1 (0.3) mT and 4 (2) $\pm$ 0.1 (0.02) mT for the 0.4 (0.5) nm Ru thickness sample,  with $M_s $= 620 (600) kA/m  for the 0.4 (0.5) nm Ru sample. The effective magnetic exchange field is obtained by differentiating the exchange coupling terms ($F_\text{ex}$) in Eq. \ref{eq:freeE} with respect to $\boldsymbol m_{1(2)} $ which is given as:
\begin{equation}
 B_\text{ex,1(2)}=-\frac{1}{2M_\text{s}}\frac{\partial F_\text{ex}}{\partial m_{1(2)}}=-\frac{J_\text{ex1}}{d}\boldsymbol{m_{2(1)}}-\frac{2J_\text{ex2}}{d}\left(\boldsymbol{m_1}\cdot\boldsymbol{m_2}\right)\boldsymbol m_{2(1)}.
\end{equation}
 \begin{figure}[h]
\centering
{\includegraphics[width=0.8\textwidth]{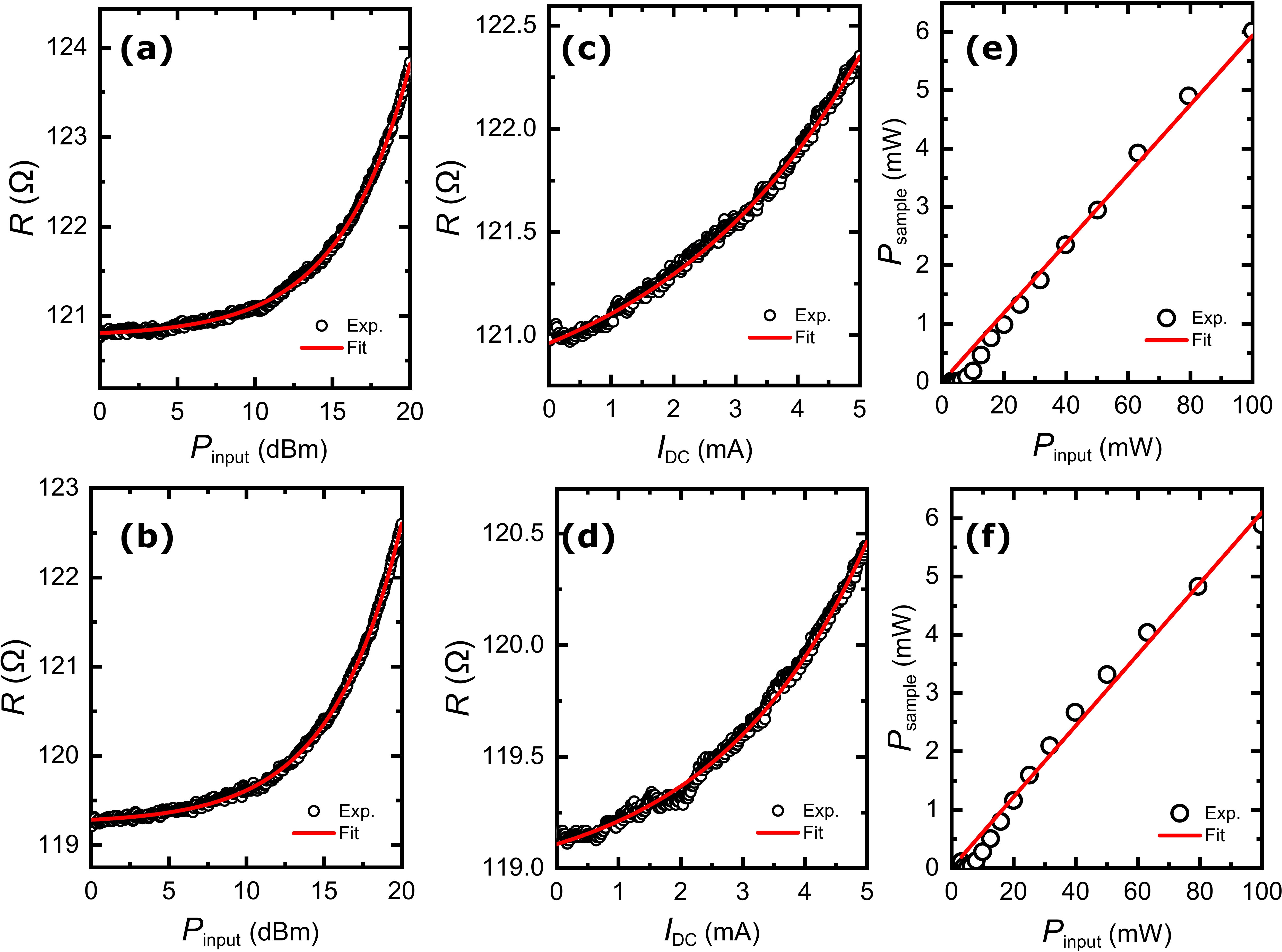}}
\caption{\label{fig:fig_4} Resistance change as a function of microwave power for (a) 0.4 nm and (b) 0.5 nm.  Resistance change as a function of dc current for (e) 0.4 nm and (f) 0.5 nm. Power from the source as a function of power in the sample for Ru thickness (e) 0.4 nm and (f) 0.5 nm.}
\end{figure}
\section{Microwave calibration}
Due to impedance mismatch between microwave lines and the sample (with few hundreds $\Omega$ in resistance), we expect that there is a large amount of power reflection from our devices. This results in that the amount of power reaching to the sample is a fraction of power supplied from the source. In order to quantify the actual power propagating through our device, we use a bolometric technique \cite{Fangnature2011,Kurebayashinature2014}. In this method, we compare the resistance change caused by joule heating from either a known dc current $I_\text{DC}$ or microwave power $P_\text{input}$ as current calibration. In Figs. \ref{fig:fig_4}(a)-(d), we show resistance changes by two current excitations. By scaling two parameters ($P_\text{input}$ and $I_\text{DC}$) by the sample resistance, we quantify the current flowing through the device at GHz frequency. We then calculate the microwave power at sample ($P_\text{sample}$) and plot it against microwave power at source ($P_\text{input}$) as shown in Figs. \ref{fig:fig_4}(e) and (f). From this slope, we extract the ratio of 0.06 between $P_\text{sample}$ and $P_\text{input}$, which would be a right value when we consider the refection from the device and microwave loss in lines and contacts in our microwave circuit.
\section{Additional STT-FMR results in this study}
\begin{figure}[h]
\centering
{\includegraphics[width=0.8\textwidth]{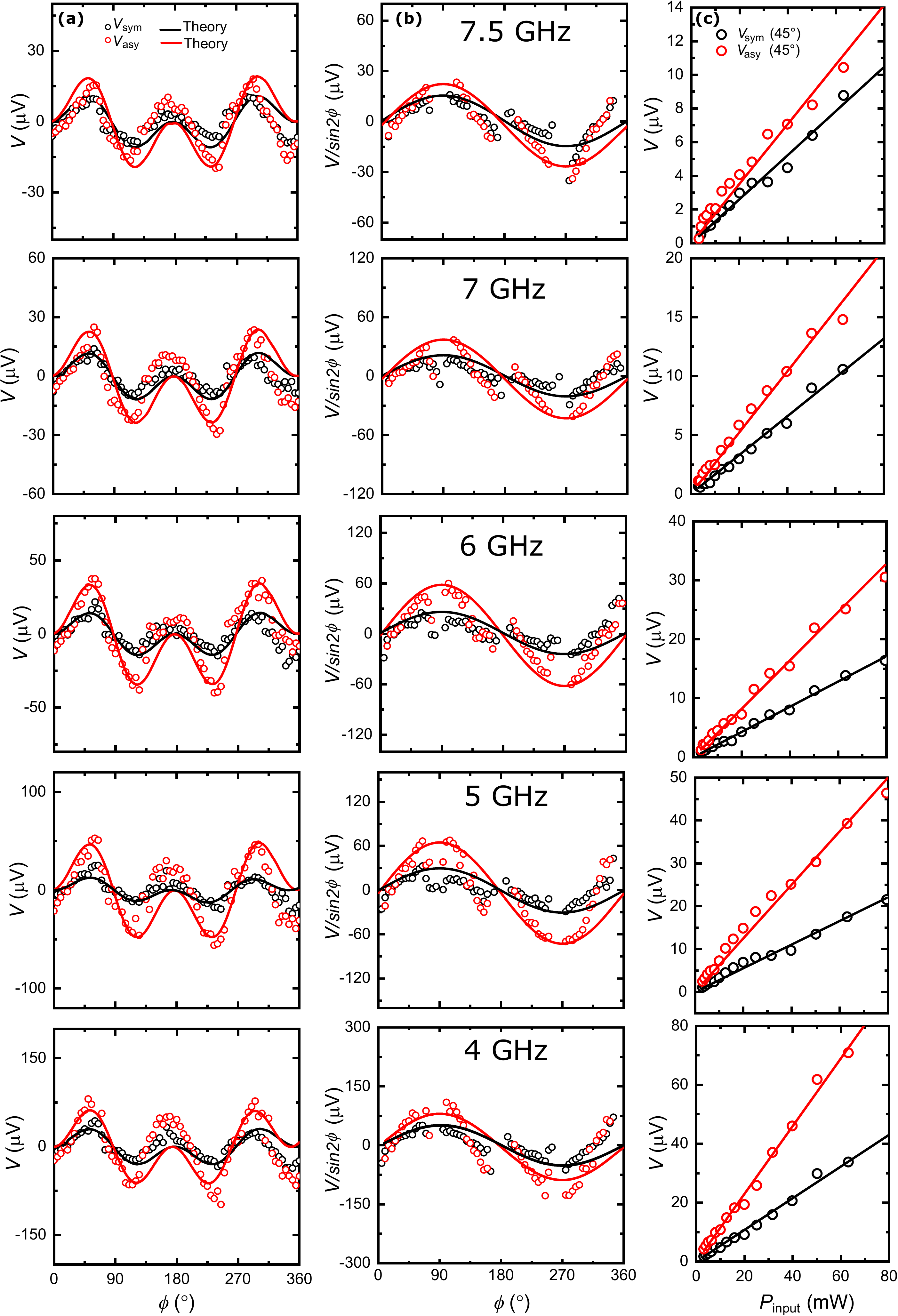}}
\caption{\label{fig:fig_5}Angular dependence of $V_\text{sym}$ and $V_\text{asy}$ for the acoustic mode measured at different frequencies from 7.5 GHz (top) to 4 GHz (bottom) as labelled in Figs. (b). Solid curves are best fit curves using our model. (b) The symmetry of torques obtained by dividing the Voltage by $\sin2\phi$. The dominant $\sin\phi$ dependence of torques confirms the parallel pumping configuration. 
(c) $V_\text{sym}$ and $V_\text{asy}$ as a function of $P$ for $\phi = 45^\circ$. The solid lines are linear fit to the data.}
\end{figure}
\begin{figure}[h]
\centering
{\includegraphics[width=0.8\textwidth]{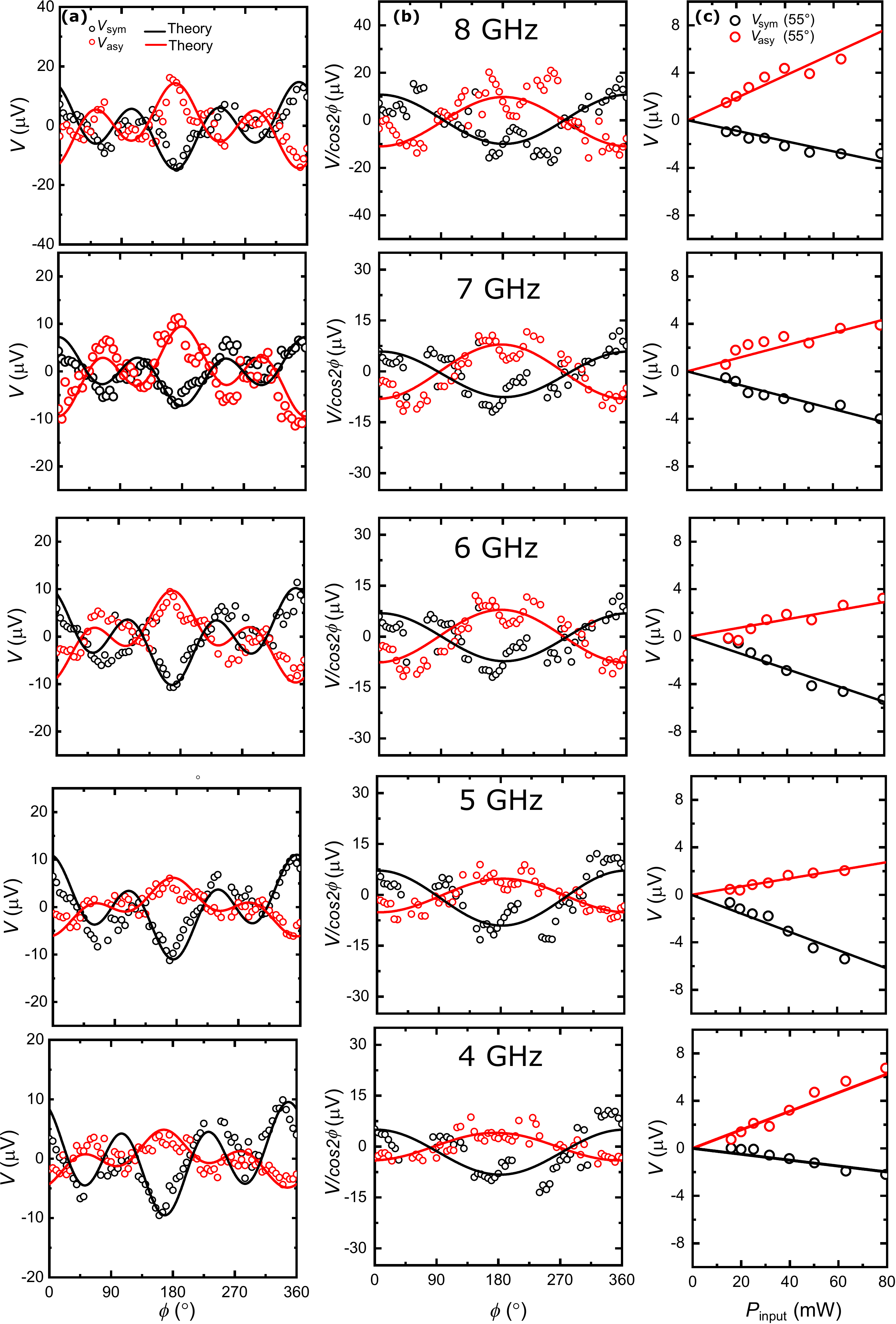}}
\caption{\label{fig:fig_6}Angular dependence of $V_\text{sym}$ and $V_\text{asy}$ for the optical mode measured at different frequencies from 8 GHz (top) to 4 GHz (bottom) as labelled in Figs. (b). Solid curves are best fit curves using our model. (b) The symmetry of torques obtained by dividing the Voltage by $\cos2\phi$. The dominant $\cos\phi$ dependence of torques confirms the perpendicular pumping configuration. (c) $V_\text{sym}$ and $V_\text{asy}$ as a function of $P$ for $\phi = 55^\circ$. The solid lines are linear fit to the data.}
\end{figure}
\begin{figure}[h]
\centering
{\includegraphics[width=1\textwidth]{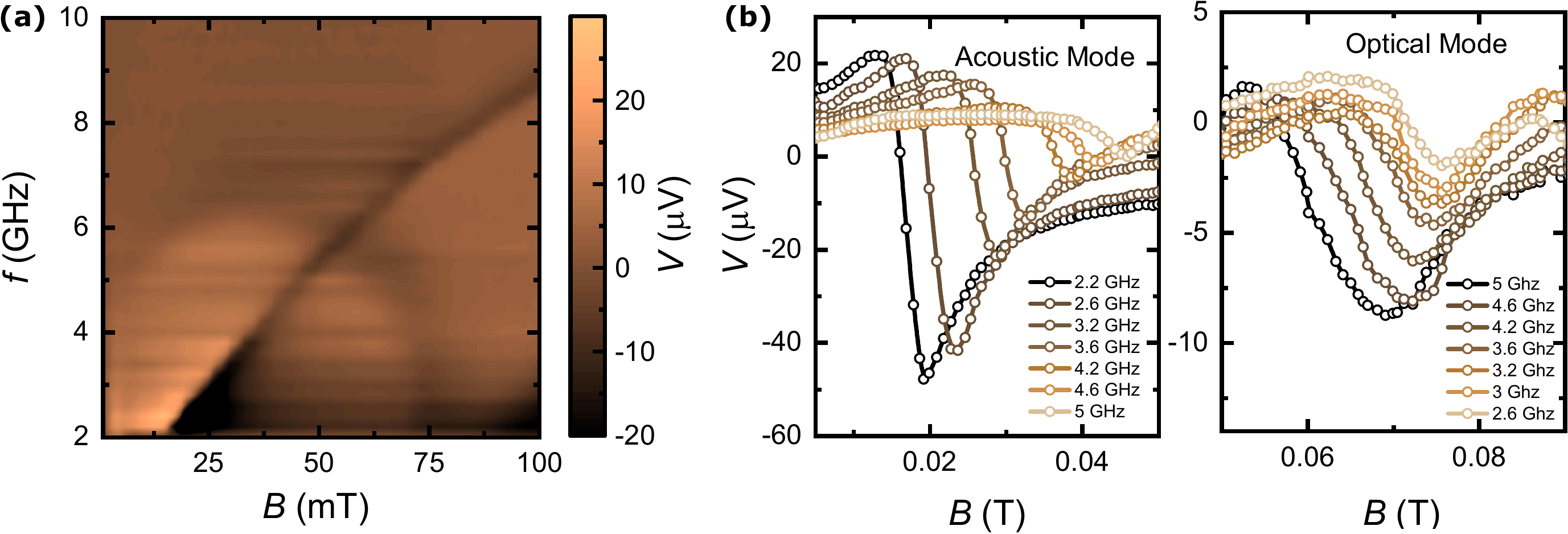}}
\caption{\label{fig:fig_7} (a) A 2D colorplot of $V$ as a function of applied field and frequency, measured for $\phi = 55^\circ $  for sample with Ru thickness 0.5 nm (b) $V$ obtained at different frequencies for acoustic and optical modes in our device for $\phi = 145^\circ$ (the values for some frequencies have been scaled to show them properly).}
\end{figure}
\begin{figure}[h]
\centering
{\includegraphics[width=0.9\textwidth]{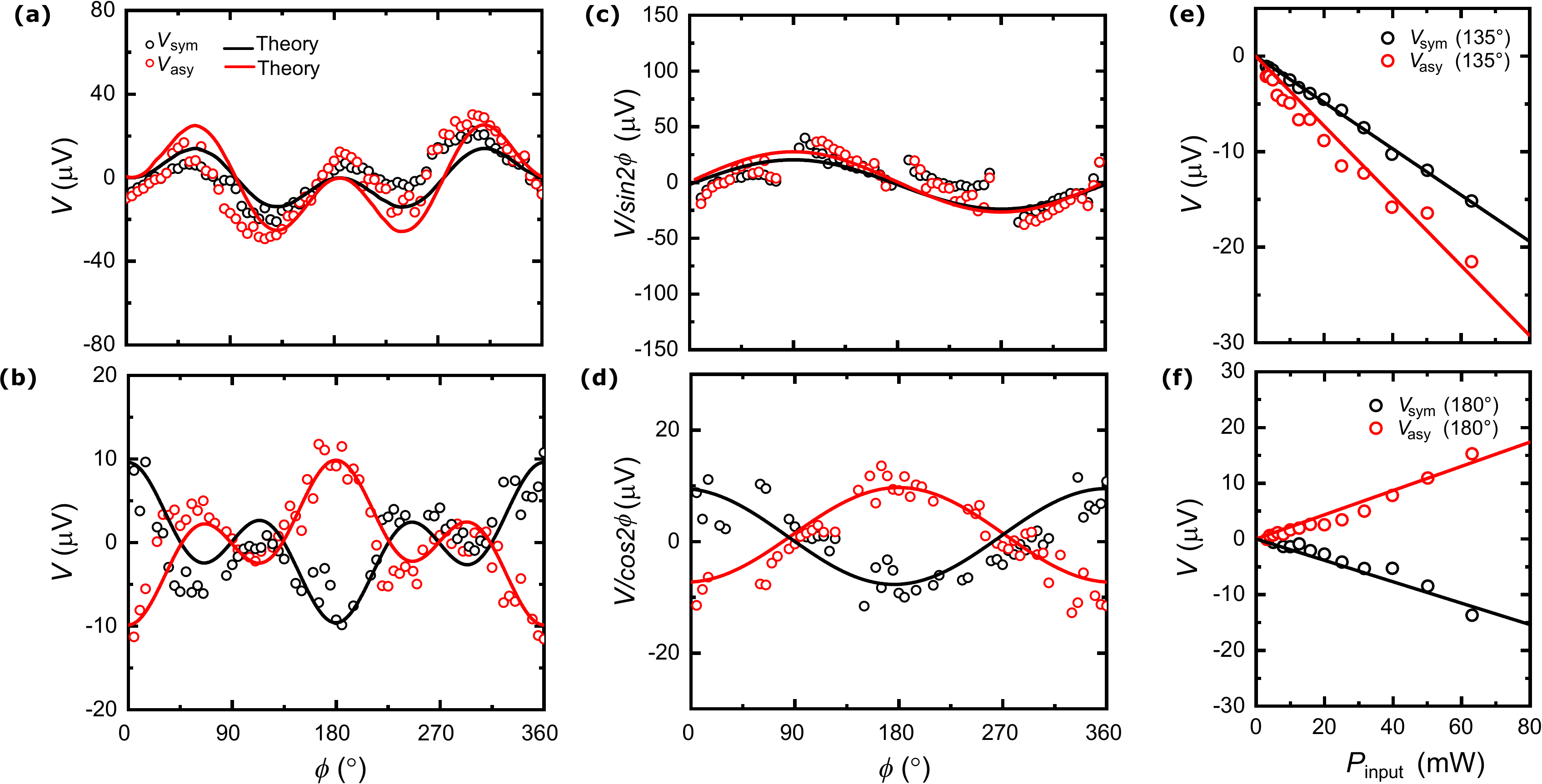}}
\caption{\label{fig:fig_8}(a-b) Angular dependence of $V_\text{sym}$ and $V_\text{asy}$ measured at a frequency of 4 GHz for Ru thickness 0.5nm for (a) Acoustic mode and (b) Optical mode . Solid lines are the  fitting curves using our model. (c-d) The symmetry of torques obtained by dividing the Voltage (c) by  $\sin2\phi$ and  (d) by $\cos2\phi$ . The parallel(perpendicular)pumping configuration is confirmed by the  dominant $\sin\phi(\cos\phi)$ dependence of torques. (e-f) $V_\text{sym}$ and $V_\text{asy}$ as a function of  input $P$ at a frequency of 4 GHz for (e) Acoustic mode at $\phi = 135^\circ$  and (f) Optical mode at $\phi = 180^\circ$. }
\end{figure}
\begin{figure}[h]
\centering
{\includegraphics[width=0.4\textwidth]{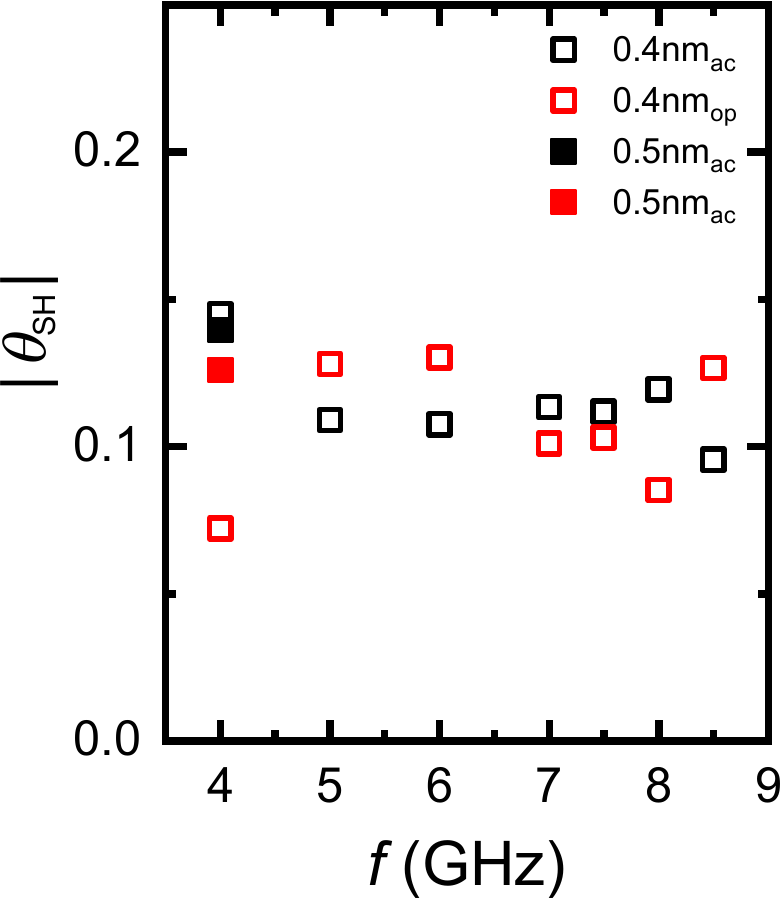}}
\caption{\label{fig:fig_11} The magnitude of spin-Hall angle $\vert\theta_{\rm SH}\vert$ in the Ta layer extracted for different frequencies with the two samples.}
\end{figure}

This section provides supplementary results used in our study to support our claims in the main text. In Fig.~\ref{fig:fig_5} we show the angular dependence of voltages for acoustic mode, the resultant torque symmetry plots and power dependence at different frequencies for the sample with Ru thickness of 0.4 nm. This supplements Fig. 2 in the main text and further supports our observation of the parallel pumping nature of acoustic mode. Similarly results for the optical mode at different frequencies are presented in Fig.~\ref{fig:fig_6}, which shows consistency of our claims across the frequency region we measured. We performed the same set of experiments on the sample with Ru thickness of 0.5 nm where the interlayer exchange field is slightly weaker. As summarised in Figures~\ref{fig:fig_8} and we show the spin hall angle calculated at different frequencies in Figure~\ref{fig:fig_11} using the procedure described in main text. We repeated similar measurements for the sample with Ru thickness of 0.5 nm.  Applying the same analysis procedure we show the resultant plots for experimental data  along with theoretical derived curves  in  Figure~\ref{fig:fig_7} and \ref{fig:fig_8}, we observe very similar results as those from the sample with Ru thickness of 0.5 nm. This further supports the validity of our claims. Finally, we show the magnitude of $\theta_{\rm SH}$ of the Ta layer extracted from our measurements for both samples and different frequencies. As already discussed in the main, the size of these values is in good agreement with that measured in previous studies\cite{LiuScience2012,SagastaPRB2018} and suggests that $V_\text{sym}$ is produced by SHE in the Ta layers in our study.  

\end{document}